\documentclass{article}
\usepackage{times}
\usepackage[utf8]{inputenc}
\usepackage{authblk}
\usepackage{amsmath}
\usepackage{amsfonts}
\usepackage{anyfontsize}
\usepackage[a4paper]{geometry}
\usepackage{enumerate}
\usepackage{setspace}
\usepackage[printonlyused]{acronym}
\usepackage[pdfpagelabels=true]{hyperref}
\usepackage{tikz}
\usepackage{multirow}
\usepackage{etoolbox}
\usepackage{natbib}
\usepackage{graphicx,rotating,booktabs}
\usepackage[verbose]{placeins}
\usepackage{blindtext}

\allowdisplaybreaks

\makeatletter
\patchcmd{\@makecaption}
  {\parbox}
  {\advance\@tempdima-\fontdimen2} 
  {}{}
\makeatother 
\newcommand{\FWER}{\textit{FWER}}
\newcommand{\FWERc}{\textit{F}}
\newcommand{\PWER}{\textit{PWER}}
\newcommand{\PWERc}{\textit{P}}

\newcommand{\mC}{\mathcal{C}}
\newcommand{\ttheta}{\tilde{\theta}}
\newcommand{\htheta}{\hat{\theta}}
\newcommand{\Pop}{\mathcal{P}}
\newcommand{\Str}{\mathcal{S}}
\newcommand{\Hd}[1]{H^{\delta_{#1}}_{#1}}
\newcommand{\Td}[1]{T^{\delta_{#1}}_{#1}}
\DeclareMathOperator*{\argmax}{arg\,max}

\colorlet{circle edge}{blue!50}
\colorlet{circle area}{blue!20}

\tikzset{filled/.style={fill=circle area, draw=circle edge, thick},
    outline/.style={draw=circle edge, thick}}

\title{A liberal type I error rate for studies in precision medicine}
\author{Werner Brannath$^1$}
 \author{Charlie Hillner$^2$} 
\author{Kornelius Rohmeyer$^3$}
\affil{$^1$University of Bremen, Institute for Statistics and Competence Center for Clinical Trials, Bremen, Germany, brannath@uni-bremen.de}
\affil{$^2$University of Bremen, Institute for Statistics and Competence Center for Clinical Trials, Bremen, Germany}
\affil{$^3$University of Oldenburg, Institute of Mathematics, Germany}
\date{\vspace{-5ex}}
\begin{document}
\maketitle
\begin{abstract}
We introduce a new multiple type I error criterion for clinical trials with multiple populations. Such trials are of interest in precision medicine where the goal is to develop treatments that are targeted to specific sub-populations defined by genetic and/or clinical biomarkers. The new criterion is based on the observation that not all type I errors are relevant to all patients in the overall population. If disjoint sub-populations are considered, no multiplicity adjustment appears necessary, since a claim in one sub-population does not affect patients in the other ones. For intersecting sub-populations we suggest to control the average multiple type error rate, i.e.\ the probability that a randomly selected patient will be exposed to an inefficient treatment. We call this the {\em population-wise error rate}, exemplify it by a number of examples and illustrate how to control it with an adjustment of critical boundaries or adjusted p-values. We furthermore define corresponding simultaneous confidence intervals. We finally illustrate the power gain achieved by passing from family-wise to population-wise error rate control with two simple  examples and a recently suggest multiple testing approach for umbrella trials.  
\end{abstract}
\section{Introduction}
The aim of precision medicine is to provide each patient with an optimal treatment tailored to his or her genetic and/or clinical profile. One strategy for reaching this goal is to undertake trials where one or several treatments are investigated in multiple sub-populations. Examples for such trials are umbrella and basket trials in oncology.
In an umbrella trial patients with the same cancer type but different molecular alterations are enrolled and the treatments are tailored to the specific target sub-populations. In a basket trial patients with different cancer types but one common molecular alteration are enrolled with the aim to study one targeted treatment 
(see e.g.\ Woodcock and LaVange, 2017; Strzebonska and Waligora, 2019).
In many cases the target or sub-populations are disjoint by nature, but when many different biomarkers or cancer types are used, it can also occur that patients belong to more than one sub-population. For example, in the FOCUS4 study (Kaplan \textit{et al.}, 2013) biomarker tests were conducted to define subgroups based on the mutations present in the patients' tumour DNA. Some patients belonged to more than one subgroup and thus the subgroups were made disjoint by means of a hierarchical ordering structure defined for the different mutations. In this manuscript, we explicitly allow biomarker-defined sub-populations to be overlapping such that patients become eligible for multiple targeted therapies. This means that future patients of the overlap may be exposed to more than a single inefficient treatment by the trial results. Moreover, for such studies suitable allocation procedures have to be defined. Issues of eligibility for multiple target therapies have been addressed e.g.\ in Malik \textit{et al.} (2014), Collignon \textit{et al.} (2020) and Kesselmeier \textit{et al.} (2020).

\nocite{doi:10.1056/NEJMra1510062,strzewali,kaplan,glimm,fletcher,MALIK20141443,Collignon,10.1371/journal.pone.0237441}

In confirmatory clinical trials with tests of several hypotheses the multiple type I error is usually kept small by controlling the family-wise error rate (FWER). With the growing effort of detecting new and more predictive biomarkers and an increasing focus on rare diseases, it is becoming more and more difficult to undertake clinical trials that are sufficiently powered and also provide sufficient control of type I errors. Since the control of multiple type I errors amplifies this issue, more liberal alternatives to the common approach of family-wise error rate control are of strong interest. 
If a treatment or a treatment strategy is tested in several disjoint populations and each population is affected by only a single hypothesis test, the overall study basically consists of separate trials that merely share the same infrastructure. Therefore, no multiplicity adjustments are needed (e.g.\ Glimm and Di Scala, 2015; Collignon 
\textit{et al.}, 2020).
However, if some sub-populations are overlapping, these intersections will contain patients that are possibly exposed to multiple erroneously rejected null hypotheses, implying that one has to adjust for multiplicity 
(e.g.\ Collignon \textit{et al.}, 2020). Since only patients in the intersections are concerned with this multiplicity issue, there is no need for adjustments for patients in the complements, who can only be affected by at most one false rejection of a null hypothesis. The FWER would therefore be too conservative also in this case. Especially for small and/or highly stratified populations, as for instance encountered in paediatric oncology, a more liberal approach is desirable (e.g.\ Fletcher 
\textit{et al.}, 2018). The purpose of this manuscript is to propose a new concept of multiple type I error control that is less conservative. With this new error rate, which we name \emph{population-wise error rate} (PWER), we aim to keep the average multiple type I error rate at a reasonable level. This provides control of the probability that a randomly chosen future patient will be exposed to an inefficient treatment policy.

The paper is outlined as follows. First, the PWER is motivated by means of a simple example, followed by the general mathematical definition. Then, we demonstrate how to control the PWER at a pre-specified level by adjusting critical boundaries or p-values. In the subsequent section the gain in power 
by using PWER instead of FWER control is illustrated by two examples. In the first example we will investigate the case of two overlapping populations and with (i) two different treatments in each population and (ii) the same treatment in both populations. The second example consists of an application of the PWER to a multiple testing approach for umbrella trials suggested in Sun \textit{et al.} (2016). In Section~\ref{sec: SCI} we extend the multiple test with PWER-control to simultaneous confidence intervals and discuss their coverage properties. The paper concludes with a discussion in Section~\ref{sec: disc}. \textcolor{black}{All computations and simulations are done in R. The corresponding R-script files are all available from the authors.}

\nocite{Sun}

\section{The population-wise error rate}\label{sec PWER}
In this section the aforementioned population-wise error rate is introduced conceptually and formally. Examples for different settings are given to further deepen the understanding. 
\subsection{General framework and definition}\label{sec: gfwd}
Consider an overall population $\Pop$ consisting of $m\ge 2$ possibly overlapping sub-populations $\Pop_1,\dots,\Pop_m$ $\subseteq \Pop$ and suppose that we want to investigate a treatment $T_i$ in each $\Pop_i$ with a sample from $\Pop_i$ and a hypothesis test. In the sequel we call the tuples $(\Pop_i, T_i)$ the \emph{treatment policies}. To each treatment policy ($\Pop_i, T_i$) we assign the null hypothesis $H_i: \theta_i \le 0$, where $\theta_i = \theta(\Pop_i, T_i)$ quantifies the efficacy of treatment $T_i$ in comparison to a control treatment in population $\Pop_i$. \textcolor{black}{We focus here on one-sided null hypotheses to avoid directional errors which are common in multiple testing.} The \emph{population-wise error rate} is then given by the risk for a randomly chosen patient to be assigned to one or more inefficient treatment policies, i.e.\ to belong to at least one tuple $(\Pop_i, T_i)$ \textcolor{black}{where for the true effect $\theta_i \le 0$ but $H_i$ has been rejected.} 

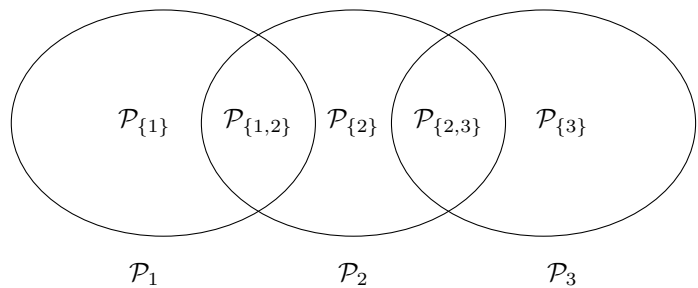
\begin{figure}[h]
\begin{center}
\begin{tikzpicture}
  \draw (0cm,0cm) ellipse[x radius=2cm,y radius=1.5cm] node at (-.25,0) {$\Pop_{\{1\}}$};
  \draw (2.5cm,0cm) ellipse[x radius=2cm,y radius=1.5cm] node{$\Pop_{\{2\}}$};
  \draw (5cm,0cm) ellipse[x radius=2cm,y radius=1.5cm] node at (5.25,0){$\Pop_{\{3\}}$};
  \node at (1.25,0) {$\Pop_{\{1,2\}}$};
  \node at (3.75,0) {$\Pop_{\{2,3\}}$};
\node at (-.25, -2) {$\Pop_1$};
\node at (2.5, -2) {$\Pop_2$};
\node at (5.25, -2) {$\Pop_3$};
\end{tikzpicture}
\end{center}
\caption{$m=3$ populations and their disjoint sub-populations}
\label{fig: 3pop2int1}
\end{figure}

In order to define the PWER mathematically, we partition the overall population into disjoint strata $\Pop_J := \bigcap_{j\in J}\Pop_j \setminus \bigcup_{k\in I\setminus J}\Pop_k$ for $J\subseteq I:=\{1,\dots,m\}$. In Figure \ref{fig: 3pop2int1} we see an example for such a partition based on three sub-populations $\Pop_i$, $i=1,2,3$. Note that $\Pop_{\{1,2,3\}}=\emptyset$. For each non-empty stratum $\Pop_J$, we denote its \textcolor{black}{proportion within the overall patient population $\Pop$ by} $\pi_J$ such that $\sum_{J\subseteq I,\, \Pop_J\not=\emptyset} \pi_J = 1$. \textcolor{black}{In most parts of the paper we will assume that the {\em relative prevalences} $\pi_J$ are known. In practical cases they will have to be estimated. The effect of estimation will be discussed in Section~\ref{sec: eopp}.} For any future patient in $\Pop_J$, $J\subseteq I$, we commit a type I error if he/she belongs to at least one $(\Pop_i, T_i)$ with $i\in J$ and $\theta_i \le 0$ for which $H_i$ has been rejected. The population-wise error rate (PWER) is then defined as 
\begin{align}\label{eq: PWER}
\PWER = \sum_{J\subseteq I,\, \Pop_J\not=\emptyset}\,\pi_J\, 
\mathbb{P}(\text{\,falsely reject any $H_i$ with $i\in J$\,}).
\end{align}
To determine the PWER, we need to know for each stratum $\Pop_J$ the probability of rejecting at least one true null hypothesis that affects the stratum. 

Compared to the FWER, which controls the maximum risk for future patients to be assigned to an inefficient treatment strategy, the PWER is an average risk. It is more liberal, because 
\begin{align*}
\PWER &= \sum_{J\subseteq I,\, \Pop_J\not=\emptyset}\,\pi_J\, 
\mathbb{P}(\text{\,falsely reject any $H_j$ with $j\in J$\,})\\
&<\left(\sum_{J\subseteq I,\, \Pop_J\not=\emptyset}\pi_J\right)\, 
\mathbb{P}(\text{falsely reject any $H_i$ for $i\in I$}) 
= \FWER.
\end{align*}
\textcolor{black}{Recall that the sum in the second last expression is equal to 1. Note that $\PWER=\FWER$ only in the here uninteresting case $\Pop=\Pop_1=\cdots=\Pop_m$.}

\subsection{Two intersecting populations}\label{sec ex1}
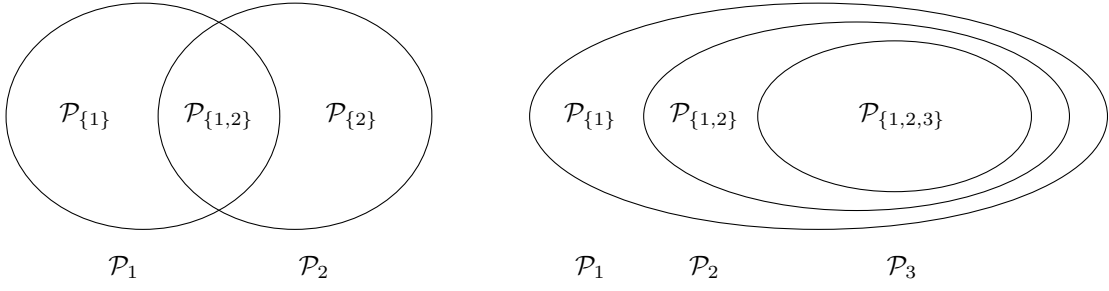
\begin{figure}[h] 
\begin{minipage}{3.8cm}
\begin{center}
\begin{tikzpicture}
  \draw (0cm,0cm) ellipse[x radius=1.8cm,y radius=1.5cm] node at (-.75,0) {$\Pop_{\{1\}}$};
  \draw (2.0cm,0cm) ellipse[x radius=1.8cm,y radius=1.5cm] node at (2.75,0) {$\Pop_{\{2\}}$};
  \node at (1.,0) {$\Pop_{\{1,2\}}$};
\node at (-.25, -2) {$\Pop_1$};
\node at (2.25, -2) {$\Pop_2$};
\end{tikzpicture}
\end{center}
\end{minipage}\hspace{3cm}
\begin{minipage}{3.8cm}
\begin{center}
\begin{tikzpicture}
  \draw (0cm,0cm) ellipse[x radius=3.8cm,y radius=1.5cm] node at (-3,0) {$\Pop_{\{1\}}$};
  \draw (.5cm,0cm) ellipse[x radius=2.8cm,y radius=1.25cm] node at (-1.5,0) {$\Pop_{\{1,2\}}$};
 \draw (1cm,0cm) ellipse[x radius=1.8cm,y radius=1cm] node at (1.1, 0) {$\Pop_{\{1,2,3\}}$};
\node at (-3, -2) {$\Pop_1$};
\node at (-1.5, -2) {$\Pop_2$};
\node at (1.1, -2) {$\Pop_3$};
\end{tikzpicture}
\end{center}
\end{minipage}
\caption{Left panel: $m=2$ intersecting populations. Right panel: $m=3$ nested populations.}
\label{fig: 3pop2int2}
\end{figure}
As an example consider a trial with two intersecting sub-populations $\Pop_1$ and $\Pop_2$ and two treatments $T_1$, $T_2$ to be tested by means of the hypotheses $H_1: \theta(\Pop_1, T_1) \le 0$
and $H_2: \theta(\Pop_2, T_2) \le 0$. Usually, the two treatments will be compared to the same control, however, the basic idea given in the subsequent sections also applies with treatment-specific controls.
As illustrated in the left panel of Fig.\ \ref{fig: 3pop2int2}, the overall population can be partitioned into three disjoint sub-populations, $\Pop_{\{1\}} := \Pop_1\setminus\Pop_2$, $\Pop_{\{2\}} := \Pop_2\setminus\Pop_1$ and $\Pop_{\{1,2\}}:= \Pop_1\cap\Pop_2$. Obviously, we commit a type I error for $\Pop_{\{i\}}$ whenever $H_i$ is falsely rejected, $i=1,2$, and for $\Pop_{\{1,2\}}$ whenever $H_1$ or $H_2$ are falsely rejected. Hence, if $H_1$ and $H_2$ are both true, then
\begin{align}\label{eq: ex1PWERboth}
\PWER = \pi_{\{1\}}\mathbb{P}(\text{reject $H_1$}) +  \pi_{\{2\}}\mathbb{P}(\text{reject $H_2$}) + \pi_{\{1,2\}}\mathbb{P}(\text{reject $H_1$ or $H_2$})\, .
\end{align}
If $H_1$ is true and $H_2$ is false, then
\begin{align*}
\PWER = \pi_{\{1\}}\mathbb{P}(\text{reject $H_1$}) + \pi_{\{1,2\}}\mathbb{P}(\text{reject $H_1$}) = 
\left(\pi_{\{1\}}+\pi_{\{1,2\}}\right)\mathbb{P}(\text{reject $H_1$})
. \end{align*}
Hence, if only one null hypothesis is true, say $H_i$, the PWER reduces to the probability of rejecting $H_i$ multiplied by the relative size of the population $\Pop_i$.

\subsection{Nested populations}
In practice, one often faces the problem of nested populations  $\Pop_1\supset \Pop_2 \supset \dots \supset \Pop_m$, as in the right panel of Fig.\ \ref{fig: 3pop2int2}.
Define the strata $\Pop_{[i]}:=\Pop_{\{1,\dots,i\}}$ for $i \le m$, \textcolor{black}{which are $\Pop_{[i]} = \Pop_{i}\setminus\Pop_{i+1}$ for $i<m$ and $\Pop_{[i]} = \Pop_{i}$ for $i=m$}. We commit a type I error for $\Pop_{[i]}$ whenever any true $H_j$ is rejected for $j\le i$. With relative prevalences $\pi_{[i]} := \pi_{\{1,\dots,i\}}$ of $\Pop_{[i]}$ the PWER under the global null hypothesis is given by
\begin{align*}
\PWER = \sum_{i=1}^m\pi_{[i]}\mathbb{P}(\text{reject at least one true $H_j$ for $j\le i$}).
\end{align*}
Especially, if $\Pop_i$ is defined by a biomarker $X$, i.e. $\Pop_i = \{X > t_i\}$ for cut-off points $t_i$, $i=1,\dots,m+1$ (with $t_{m+1}:= \infty$), the PWER under global null hypothesis can be written as
\begin{align*}
\PWER = \sum_{i=1}^m\mathbb{P}(t_i < X \le t_{i+1})\mathbb{P}(\text{reject at least one true $H_j$ for $j\le i$}).
\end{align*}

\subsection{Three populations with two intersections}
Finally, we give an example where the FWER is strictly conservative even for control of the maximum (instead of the average) type I error rate. Consider three populations $\Pop_1$, $\Pop_2$, $\Pop_3$ with $\Pop_1\cap\Pop_2 \not= \emptyset$, $\Pop_2\cap\Pop_3\not= \emptyset$ and $\Pop_1\cap\Pop_3 = \emptyset$, as in Fig.\ \ref{fig: 3pop2int1}. 
Under the global null hypothesis, where all null hypotheses $H_i$ are true, the PWER is given by 
\begin{align*}
\PWER = \sum_{i=1}^3\pi_{\{i\}}\mathbb{P}(\text{reject $H_i$}) + \sum_{i=1}^2\pi_{\{i,i+1\}}\mathbb{P}(\text{reject $H_i$ or $H_{i+1}$}).
\end{align*}
The FWER under the global null hypothesis equals 
$\mathbb{P}(\text{reject $H_1$ or $H_2$ or $H_3$})$.
Since no patient can belong to $\Pop_1$ and $\Pop_3$ simultaneously, the FWER corrects for a multiplicity no patient is affected by.

\section{Control of the population-wise error rate}\label{sec_CofPWER}

In this section we demonstrate how to achieve control of the PWER at a pre-specified level $\alpha$ under the general framework of Section~\ref{sec: gfwd}. Suppose that each $H_i$ can be tested with a test statistic $Z_i$ where larger values of $Z_i$ speak against $H_i$. We assume further that the joint distribution of $\{Z_i\}_{i=1}^m$ is known (at least asymptotically). In order to control the PWER at a pre-specified significance level $\alpha\in(0,1)$, we need to find the smallest critical value $c^*\in\mathbb{R}$ such that 
\begin{align}
\label{eq: PWEReq}
\PWER_{\boldsymbol\theta^\ast} = \sum_{J\subseteq I}\pi_J\mathbb{P}_{\boldsymbol\theta^*}\left(\bigcup_{i\in J\cap I(\theta^\ast)}\{Z_i > c^*\}\right) \le \alpha, 
\end{align}
where $\boldsymbol\theta^* = (\theta_1^*,\dots, \theta_m^*)$ is the parameter configuration that maximizes the PWER and 
 $I(\boldsymbol\theta^\ast)=\{i\in I: \theta^\ast_i\in H_i\}$ is the index set of corresponding true null hypotheses. Usually the maximal PWER is obtained under the global null hypothesis, i.e.\ for $\boldsymbol\theta^* = (0,\dots,0)$. 
If the joint distribution of the $Z_i$ is continuous, then we can reject $H_i$ also if $Z_i=c^*$, i.e\ the strict inequalities in (\ref{eq: PWEReq}) can be replaced by the more familiar rules $Z_i\ge c^\ast$. 

Since the (asymptotic) correlations between the test statistics usually depend only on the relative prevalences $\pi_J$, $J\subseteq I$, the PWER-level can be exhausted under $\theta^\ast$. 
When each $H_i$ is tested by means of a p-value $p_i$, we can reach $\PWER \le \alpha$ by choice of an adjusted significance level $\alpha^*$ applied to all $p_i$. 

The critical value $c^*$ in (\ref{eq: PWEReq}) or adjusted significance level $\alpha^\ast$ can be solved by applying a univariate root finding method. Because the PWER is always bounded by the FWER, the critical value and adjusted significance level are more liberal than for FWER-control. Therefore the PWER leads to a higher power and a lower sample size to achieve a certain power. 

Instead of determining the critical value $c^\ast$ we could report the {\em PWER-adjusted} p-values 
\begin{align}
\label{eq: PWERpValue}
p^\PWER_j= \sum_{J\subseteq I}\pi_J\mathbb{P}_{\boldsymbol\theta^*}\left(\bigcup_{i\in J\cap I(\theta^\ast)}\{Z_i > z^\text{obs}_j\}\right), \quad j=1,\ldots,m,
\end{align}
where $z^\text{obs}_j$ is the observed value of $Z_j$. Obviously, $p^\PWER_j\le \alpha$ if an only if $z^\text{obs}_j\ge c^\ast$ and hence $H_j$ can alternatively be tested with the PWER-adjusted p-value $p^\PWER_j$. Furthermore, $p^\PWER_j$ gives the smallest $\PWER$-level the hypothesis $H_j$ can be rejected with. 

Note that we could control the PWER also with population-specific critical values $c^\ast_i$ (or adjusted levels $\alpha^\ast_i$). Unique solutions for $c^\ast_{i}$ can be obtained by setting $c^*_{i} = w_i c^*$ for pre-specified weights $w_i>0$ and searching for the $c^\ast$ that meets the pre-specified PWER-level. Multiplicity adjusted p-values can also be calculated with the weights $w_i$. 

The weights may, for instance, be larger for smaller populations $\Pop_i$ in order to increase the chance of finding efficient treatment policies for small sub-populations. However, due to the weighting by $\pi_J$ in definition (\ref{eq: PWER}) and expression (\ref{eq: PWEReq}), the multiple type I error rate  $\mathbb{P}_{\boldsymbol\theta^*}\left(\bigcup_{i\in J\cap I(\theta^\ast)}\{Z_i \ge c^*\}\right)$ for $\Pop_J$ will automatically be larger for smaller $\pi_J$. 
We will therefore only consider equal critical values $c^\ast_i=c^\ast$ in our examples below. 

\section{Comparison with FWER-controlling procedures}
Due to the PWER being more liberal than the FWER, the next naturally arising question is how much this affects power and sample size. We will at first compare PWER-control with FWER-control 
for two intersecting sub-populations  when (i) the same and (ii) two different treatments are investigated in each sub-population. Secondly, we will apply PWER-control to the multiple testing approach for umbrella trials considered in Sun \textit{et al.} (2016) and compare it to the originally suggested FWER-control. 

\subsection{Combination of independent studies}
We start with a hypothetical, but statistically simple situation.
Assume that a treatment $T$ is investigated for two intersecting populations $\Pop_i$, $i=1,2$, that are defined by two different biomarkers. Assume further that a sponsor has decided to test the effect of $T$ for the two biomarker positive groups in two different but parallel clinical trials with different centres. Since the two studies are submitted as a package to regulatory authorities, a multiple testing approach is required. Let us assume that PWER-control is accepted as a compromise between control of the FWER and the unadjusted testing, the latter being the case when submitting the two studies one after another. PWER-control bounds the overall probability for a future patient to be exposed to an inefficient treatment strategy.

Since the two treatment strategies $(\Pop_i,T)$, $i=1,2$, are investigated in two independent studies, the corresponding test statistics $Z_i$ are stochastically independent. Let us further assume that both $Z_i$ are normally distributed with variance 1. The question is now, what we gain in terms of power by switching from FWER- to PWER-control. We will assume an overlap between the two populations $\Pop_1$ and $\Pop_2$ of probability $\pi_{\{1,2\}}$ that will be varied in our investigation.

Let $\Phi$ and $\Phi^{-1}$ be the standard normal distribution and quantile functions, respectively. 
By the independence of the test statistics, the $\FWER=1-\Phi(c_{\FWERc}^*)^2$ 
is controlled at $\alpha$ by $\check{\text{S}}$id$\acute{\text{a}}$k's  critical value 
$c_{\FWERc}^* = \Phi^{-1}(\sqrt{1-\alpha})$. Following Example 1, the PWER is given by 
\begin{align*}
\PWER &= (1-\pi_{\{1,2\}})\left\{ 1-\Phi(c_{\PWERc}^*)\right\}+\pi_{\{1,2\}}\{1-\Phi(c_{\PWERc}^*)^2\}
\end{align*}
where $c_{\PWERc}^*$ is the critical value used for control of the PWER at level $\alpha$.
Note that $\pi_{\{1,2\}}$ determines how much multiplicity adjustment is needed for PWER-control.
Solving $\PWER = \alpha$ yields 
\begin{align}
c_{\PWERc}^* = \Phi^{-1}\left(\frac{-(1-\pi_{\{1,2\}})+ \sqrt{(1-\pi_{\{1,2\}})^2+4\pi_{\{1,2\}}(1-\alpha)}}{2\pi_{\{1,2\}}}\right);
\end{align}
see Appendix A for the derivation. For $\pi_{\{1,2\}} \downarrow 0$ this critical value decreases to $\Phi^{-1}(1-\alpha)$ coinciding with the unadjusted case and for $\pi_{\{1,2\}} \uparrow 1$ we have $c_{\PWERc}^* \uparrow c_{\FWERc}^*$.

To assess the power gain by using PWER- instead of FWER-control, we consider the factor of sample size increase with PWER- or FWER-control in comparison to the one with no multiplicity correction. Aiming for a marginal power of at least $1-\beta$, the sample size for each population $\Pop_j$ has to be at least 
$n_c \ge (\Phi^{-1}(1-\beta) + c)^2/\delta_j^2$ 
with critical value $c$ and non-centrality parameter $\delta_j$ in $\Pop_j$. The fractions
\begin{align}
\label{eq: ssincrease}
q_{\alpha}(c) := \frac{n_c}{n_{\Phi^{-1}(1-\alpha)}} = \left(\frac{\Phi^{-1}(1-\beta) + c}{\Phi^{-1}(1-\beta) + \Phi^{-1}(1-\alpha)}\right)^2 \quad \text{for $c\in\{c_{\PWERc}^*, c_{\FWERc}^*\}$},
\end{align}
describe how much more sample size one would need for a marginal power of $1-\beta$ when the multiplicity adjustments are performed.

\begin{figure}[h]
\centering
\makebox{\includegraphics[width=0.5\textwidth]{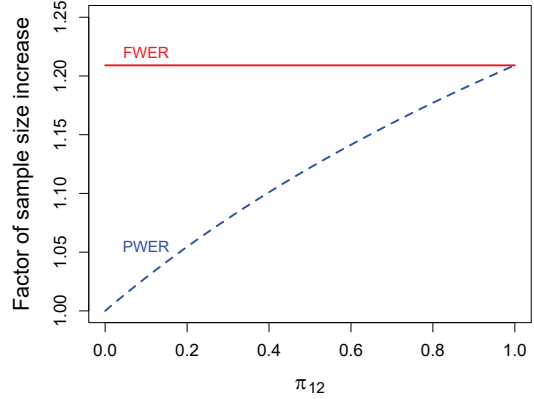}}
\caption{\label{fig: PWER_indep_plot} Factor of sample size increase compared to the unadjusted case to achieve a marginal power of $1-\beta = 80\%$ with PWER- and FWER-control at \textcolor{black}{$\alpha = 0.025$} in a combination of two independent studies with different but overlapping populations.}
\end{figure}

Figure \ref{fig: PWER_indep_plot} shows $q_{\alpha}(c)$ for $\alpha = 0.025$ \textcolor{black}{and $\beta = 0.2$} depending on the size $\pi_{\{1,2\}}$ of $\Pop_1\cap\Pop_2$ when both populations are assumed to be of equal size. FWER-control requires an increase in sample size of about 21\%  while PWER-control requires considerably less depending on $\pi_{\{1,2\}}$. The larger the intersection, the more patients are potentially exposed to two false rejections, therefore the critical value increases and the sample size increases as well.  At $\pi_{\{1,2\}} = 1$, PWER and FWER coincide and so do the sample sizes. If, for instance, the intersection makes up $40\%$ of the union of the two populations only around $10\%$ sample size increase is needed when using PWER-control, less than half of what is necessary with FWER-control. 

\subsection{Testing population specific effects in one study}\label{sec_OneStudy}

We consider now a single study with two overlapping populations $\Pop_i$, $i=1,2$, for each of which a treatment $T_i$ is compared to a common control $C$. We will investigate two possible scenarios, namely (i) $T_1 \not= T_2$ and (ii) $T_1 = T_2$. For simplicity, we assume again that both populations have the same size, i.e.\ $\pi_{\{1\}} = \pi_{\{2\}}$. We assume further that the data from each population are normally distributed with mean treatment difference $\theta_i$ and a common known variance $\sigma^2$ (across treatments and subgroups) and z-tests are used to test $H_i:\theta_i \le 0$.  For $J\subseteq \{1,2\}$, we denote by $n_J = N\cdot \pi_J$ the sample size in $\Pop_J$ and by $N = \sum_{J\subseteq \{1,2\}}n_J$ the overall total sample size.\\

\textbf{(i) Unequal treatments:} In scenario (i) we have to think of a way to randomize patients to either treatment or control. In the complements $\Pop_{\{i\}}$ we simply apply 1:1 randomization to treatment $T_i$ or control $C$. In the intersection $\Pop_{\{1,2\}}$ we apply 1:1:1 randomization to the three groups $T_1, T_2$ and $C$. By this we can assume that in $\Pop_{\{i\}}$ there are $n_{\{i\}}/2$ patients in the treatment and control group, whereas in the intersection there are $n_{\{1,2\}}/3$ patients in each group. 

Obviously, this type of allocation leads to an inconsistency between the sample sizes and prevalences. Say $\Pop_1$ has a prevalence of $\pi_1 = \pi_{\{1\}} + \pi_{\{1,2\}} = 100/170=0.59$ and of 100 patients in $\Pop_1$, 70 belong to $\Pop_{\{1\}}$ and 30 to $\Pop_{\{1,2\}}$. However, applying the above allocation rule implies that $35/45 \approx 77.7\%$ of the patients sampled from $\Pop_1$ and assigned to treatment $T_1$ belong to $\Pop_{\{1\}}$. This means that the proportions of the strata-wise sample sizes within a treatment group do not match their corresponding proportions in the population. Hence, the 
population-wise means must be estimated by a weighted sum of strata-wise means: 
\begin{align*}
\hat{x}_{i,G_i} = \left(\dfrac{\pi_{\{i\}}}{\pi_{\{i\}}+\pi_{\{1,2\}}}\right)\bar{x}_{\{i\},G_i} + \left(\dfrac{\pi_{\{1,2\}}}{\pi_{\{i\}}+\pi_{\{1,2\}}}\right)\bar{x}_{\{1,2\},G_i},\quad G_i\in\{T_i,C\},
\end{align*} 
where $\bar{x}_{J,G_i}$ is the mean response in strata $\Pop_J$, $J\subseteq \{1,2\}$, under treatment $G_i$.
In the above example, we would need to compute $\hat{x}_{1,T_1} = 0.7\cdot\bar{x}_{T_1, \{1\}} + 0.3\cdot\bar{x}_{T_1, \{1,2\}}$ for treatment $T_1$.

The z-test statistic is finally given by 
$Z_i = \left(\hat{x}_{i,T_i} - \hat{x}_{i,C}\right)/\sqrt{\text{Var}(\hat{x}_{i,T_i} - \hat{x}_{i,C})}$.
Since in the intersection $\Pop_{\{1,2\}}$ the same control group is used for both test statistics, they are positively correlated. Assuming $\pi_{\{1\}} = \pi_{\{2\}}$, 
we obtain 
$\text{Corr}(Z_1, Z_2) = (3/2)\pi_{\{1,2\}}/(1+2\pi_{\{1,2\}})$. 
The calculation of this correlation and an expression for the variance $\text{Var}(\hat{x}_{i,T_i} - \hat{x}_{i,C})$ can be found in Appendix B.\\

\textbf{(ii) Equal treatments:} In scenario (ii), we investigate one and the same treatment $T_1=T_2=T$ in both populations and apply the 1:1 randomization to every stratum. By this
 we can use for $H_i$ the test statistic
$Z_i = \left(\bar{x}_{T_i}-\bar{x}_{C}\right)/\left(2\sigma/\sqrt{n_{\{i\}}+n_{\{1,2\}}}\right)$.
Because we are using the same treatment in both populations, we expect a higher correlation between $Z_1$ and $Z_2$. Indeed, for $\pi_{\{1\}} = \pi_{\{2\}}$ the correlation  is equal to
$\text{Corr}(Z_1, Z_2) = {2\pi_{\{1,2\}}}/{\left(1+\pi_{\{1,2\}}\right)}$
which is greater or equal to the correlation with different treatments for all $\pi_{\{1,2\}}\in[0,1]$; see Appendix B.\\

For both scenarios, we intend to find critical values to control the PWER and FWER, respectively. 
Following Section~\ref{sec ex1} the PWER under the global null is given by
\begin{align}\label{eq_PWER_OnePop}\nonumber
\PWER &= \pi_{\{1\}}\mathbb{P}(\{Z_1 \ge c_{\PWERc}^*\}) +  \pi_{\{2\}}\mathbb{P}(\{Z_2 \ge c_{\PWERc}^*\})
+ \pi_{\{1,2\}}\mathbb{P}(\{Z_1 \ge c_{\PWERc}^*\}\cup \{Z_2 \ge c_{\PWERc}^*\})\\
&= (1-\pi_{\{1,2\}})\{1-\Phi(c_{\PWERc}^*)\} + \pi_{\{1,2\}}\{1-\Phi_{\rho
}(c_{\PWERc}^*,c_{\PWERc}^*)\}\end{align}
with $c_{\PWERc}^*$ being the critical value that is to be found, and $\Phi_{\rho
}$ is the cumulative distribution function of the bivariate normal distribution with standard normal marginals and correlation $\rho$. A univariate root finding algorithm can now be used to solve $\PWER = \alpha$ for $c_{\PWERc}^*$.
\begin{figure}
\centering
  \begin{minipage}[b]{0.45\textwidth}
    \makebox{\includegraphics[width=\textwidth]{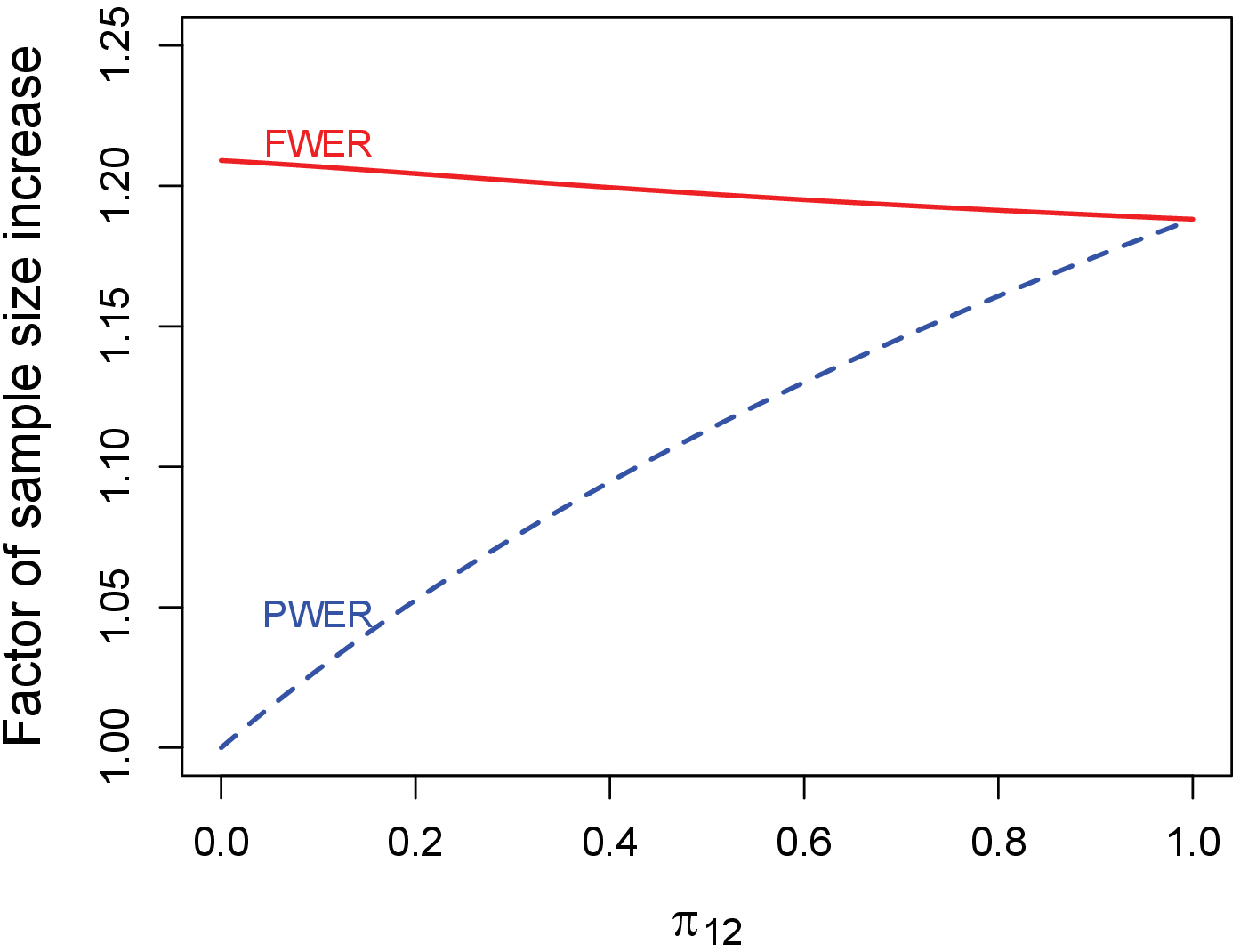}}
  \end{minipage}
  \begin{minipage}[b]{0.45\textwidth}
    \makebox{\includegraphics[width=\textwidth]{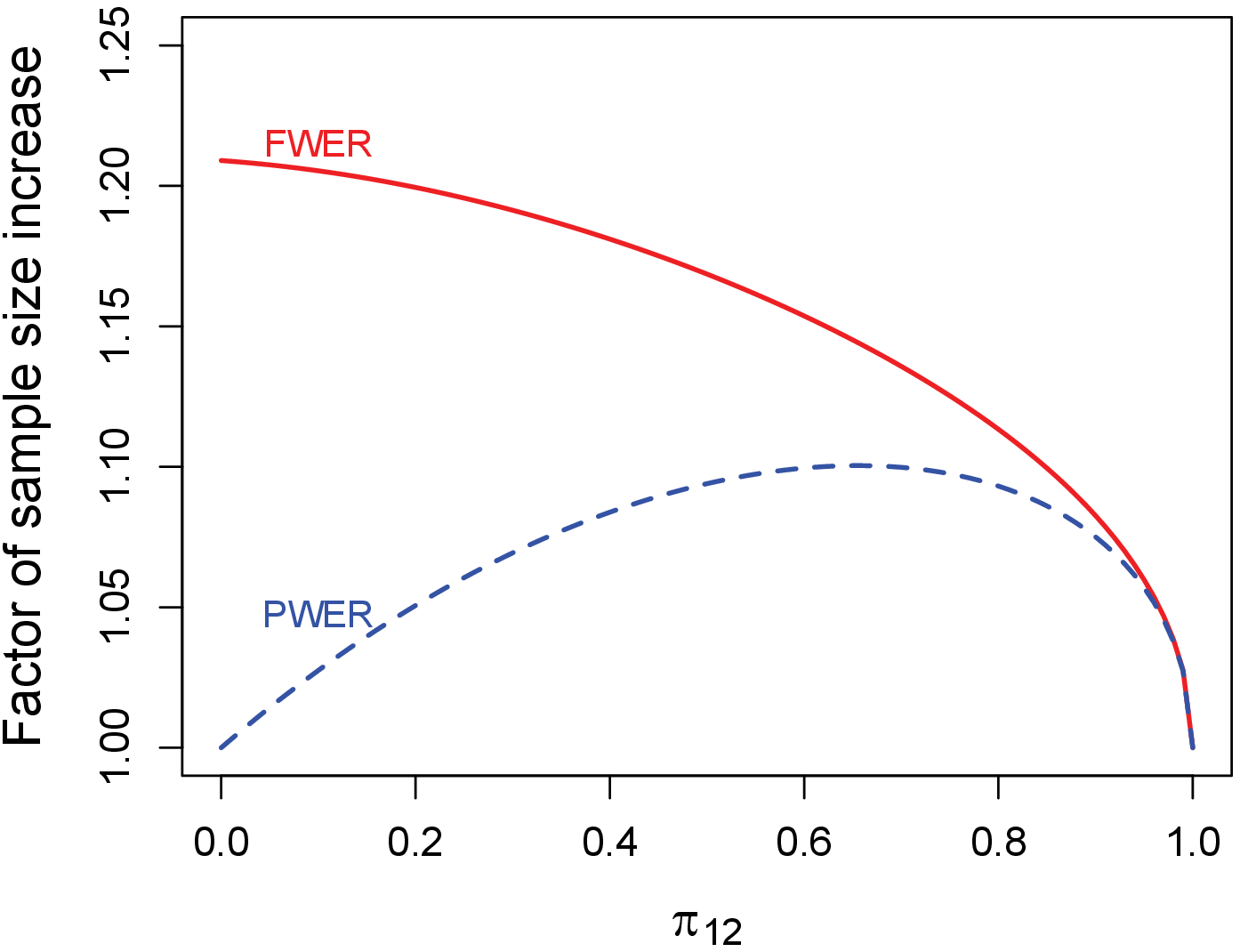}}
  \end{minipage}
\caption{\label{fig: fwervspwer} Factor of sample size increase compared to the unadjusted case for FWER- and PWER-control \textcolor{black}{at $\alpha = 0.025$} in a single study with two overlapping populations depending on the size of the intersection $\pi_{\{1,2\}}$. The Left panel is for scenario (i) with different experimental treatments and a common control; the right panel is for scenario (ii) with the equal experimental treatments. \textcolor{black}{The power is $1-\beta = 80\%$ in both scenarios.}}
\end{figure}

As an example, suppose we are in scenario (i) (multiple treatments) with $\pi_{\{1\}} = \pi_{\{2\}} = 0.4$, $\pi_{\{1,2\}} = 0.2$, $\beta = 0.2$ and $\alpha = 0.025$. Then we have $\rho=\text{Corr}(Z_1, Z_2) \approx 0.01$.
We solve $\FWER = 1-\Phi_{\rho}(c_{\FWERc}^*,c_{\FWERc}^*) = \alpha$ to obtain $c_{\FWERc}^* \approx 2.23$ and $\PWER = \alpha$ to obtain $c_{\PWERc}^* \approx 2.03$. Using (\ref{eq: ssincrease}), this yields a sample size increase of around 20\% for the FWER and only an increase of 5\% for the PWER. 

Figure \ref{fig: fwervspwer} shows graphs of sample size increases for \textcolor{black}{both scenarios} and both types of multiple error control in dependence of $\pi_{\{1,2\}}$. At $\pi_{\{1,2\}} = 0$ (disjoint populations), for instance, the PWER-approach yields no sample size increase, where the FWER-based method yields an increase of over $20\%$. With increasing intersection size the difference between sample sizes for PWER and FWER control declines until both values fall together at $\pi_{\{1,2\}} = 1$ where the PWER is equal to the FWER. 

For the PWER, this graphic also illustrates that the correlation of the test statistics and the degree of adjustments needed to correct for multiplicity behave like opposing 'forces'. At $\pi_{\{1,2\}} = 0$ the test statistics are uncorrelated, but there is no need to adjust for multiplicity with PWER-control. At $\pi_{\{1,2\}} = 1$ there is only one population, so the correlation is 1 which implies again that no multiplicity adjustment is needed, although we are formally testing two hypotheses for everyone. 
For $\pi_{\{1,2\}}$ between 0 and 1 we obtain the maximum for the PWER and corresponding sample size increase. 
Mathematically, this can be seen by rewriting the PWER as
$1-\Phi(c^*) + \pi_{\{1,2\}}\{\Phi(c^*) - \Phi_{\rho}(c^*,c^*)\}$.
For fixed $c^*$, only the second term depends on $\pi_{\{1,2\}}$. 
It is the product of two non-negative factors, $\pi_{\{1,2\}}$ 
and $\Phi(c^*) - \Phi_{\rho}(c^*,c^*)$), where the first increases from 0 to 1 and the second decreases from 1 to 0.
\subsection{Estimation of population prevalences}\label{sec: eopp}

In clinical practice the assumption of known prevalences $\pi_J$ is rarely justified and it is natural to ask whether the replacement of $\pi_J$ by an estimator $\hat{\pi}_J$ will significantly inflate the PWER. A suitable choice for $\hat{\pi}_J$ is the maximum likelihood estimator (MLE) from the multinomial distribution of $(n_J)_{J\subseteq I}$. Using these estimates instead of the true prevalences, we compute the critical value $c^\ast$ by solving $PWER = \alpha$. This guarantees asymptotic control of the PWER, since $(\hat{\pi}_J)_{J\subseteq I}$ is consistent and the joint distribution of the test statistics used in the calculation of $c^\ast$ is conditional on $(n_J)_{J\subseteq I}$.   

\begin{figure}
\noindent
  \begin{minipage}[ht]{0.47\linewidth}
    \centering
    \includegraphics[width=\linewidth]{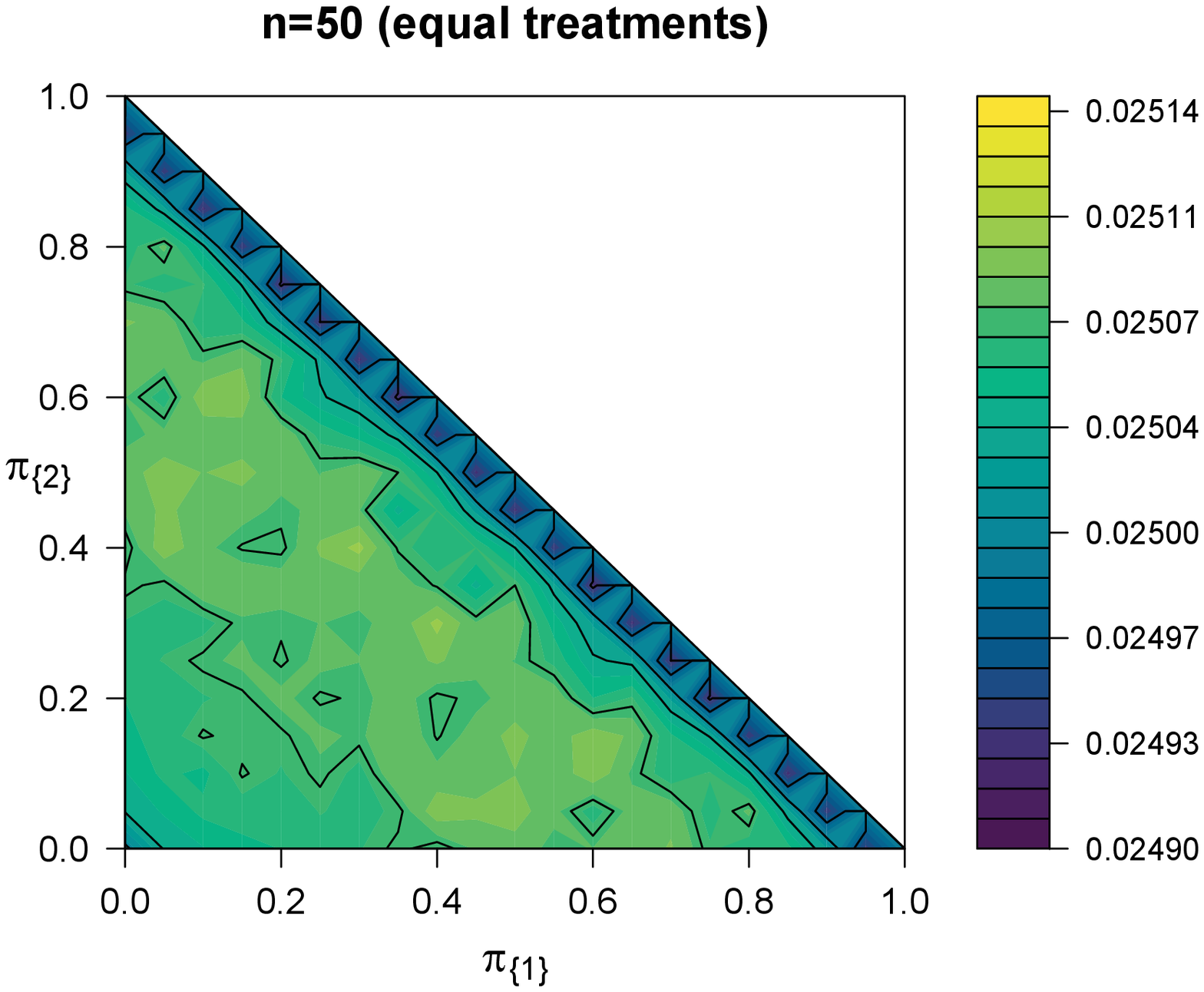} 
  \end{minipage}\hfill
  \begin{minipage}[ht]{0.47\linewidth}
    \centering
    \includegraphics[width=\linewidth]{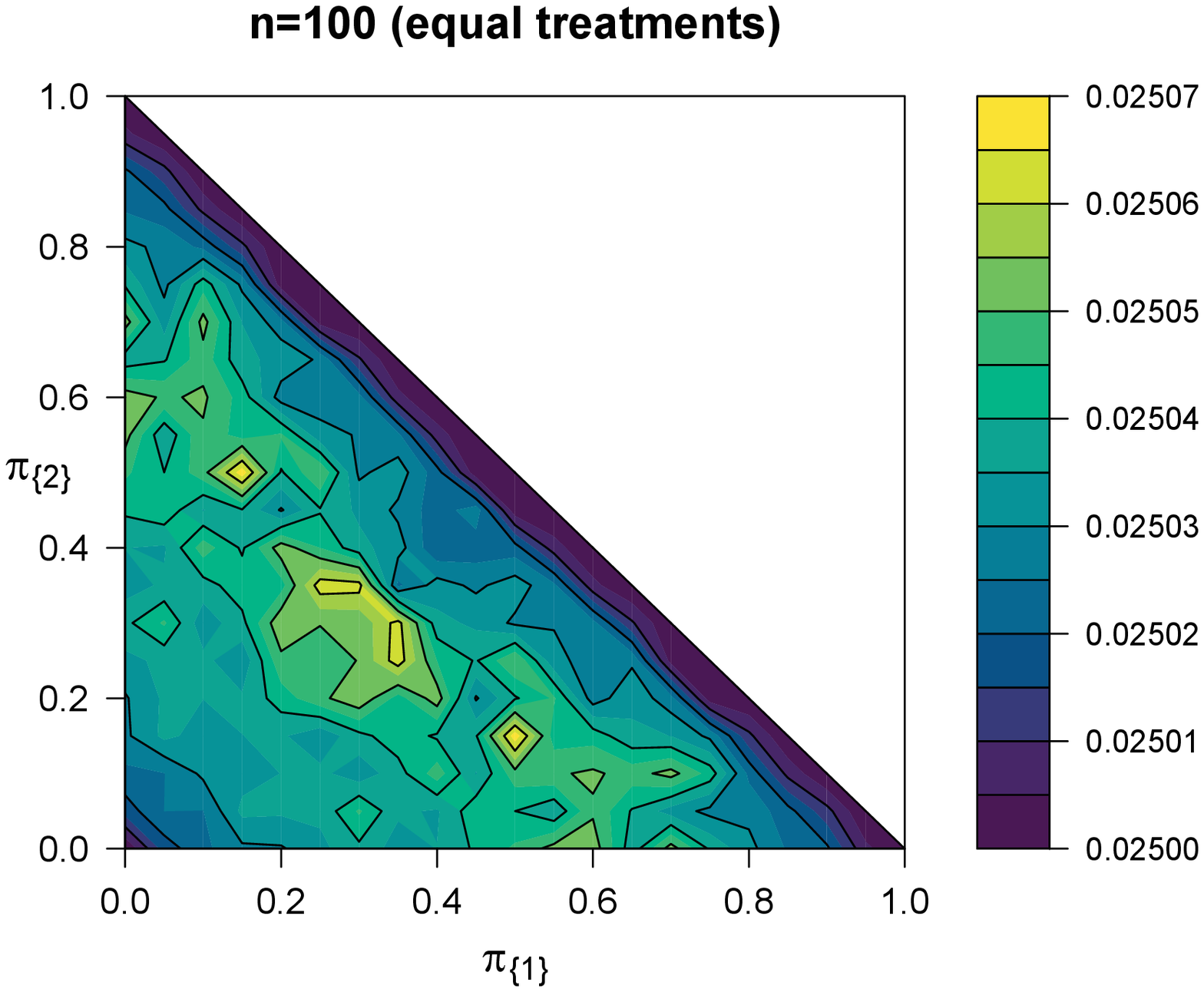} 
  \end{minipage}\hfill
  \begin{minipage}[ht]{0.47\linewidth}
    \centering
    \includegraphics[width=\linewidth]{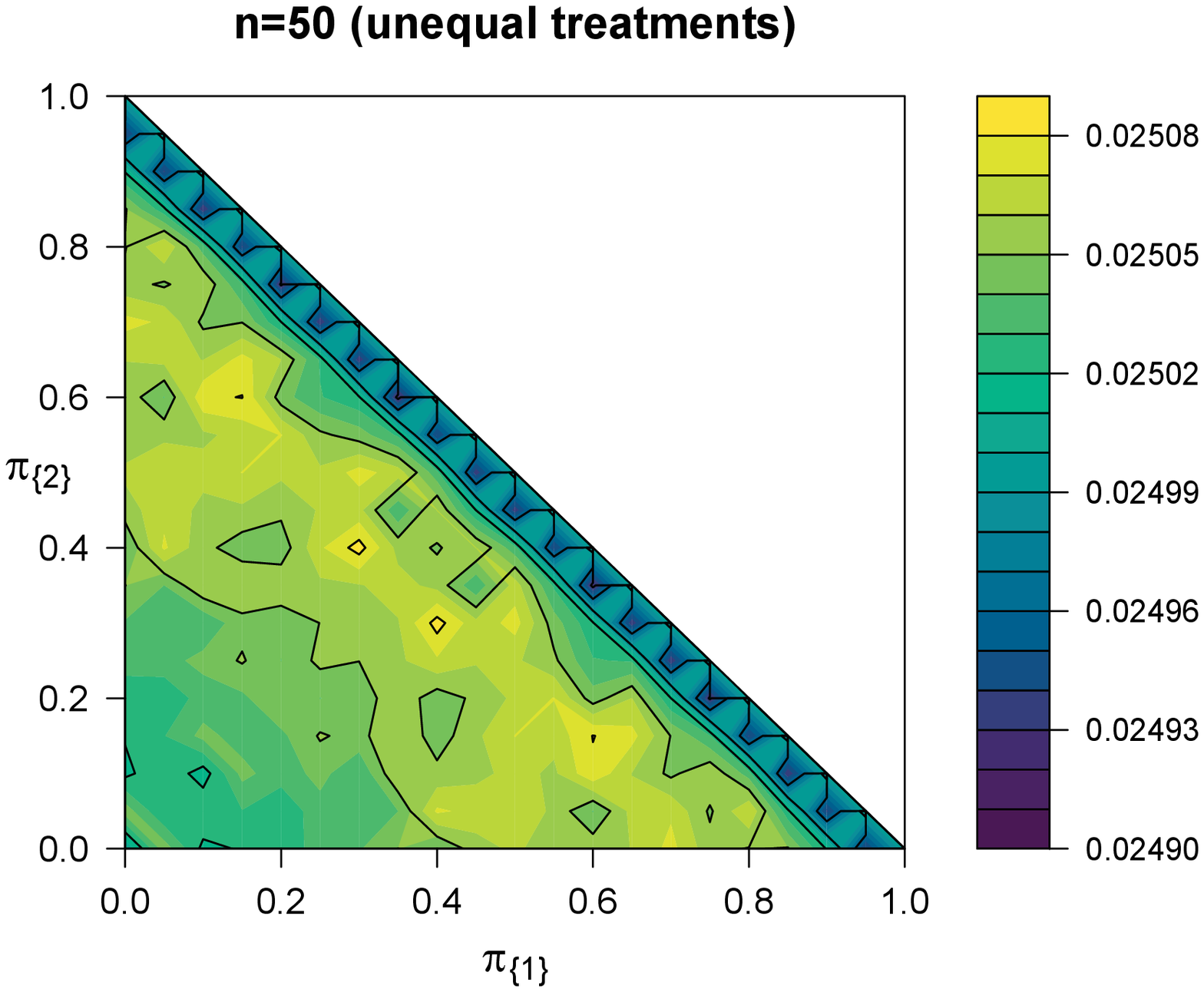} 
  \end{minipage} \hfill
  \begin{minipage}[ht]{0.47\linewidth}
    \centering
    \includegraphics[width=\linewidth]{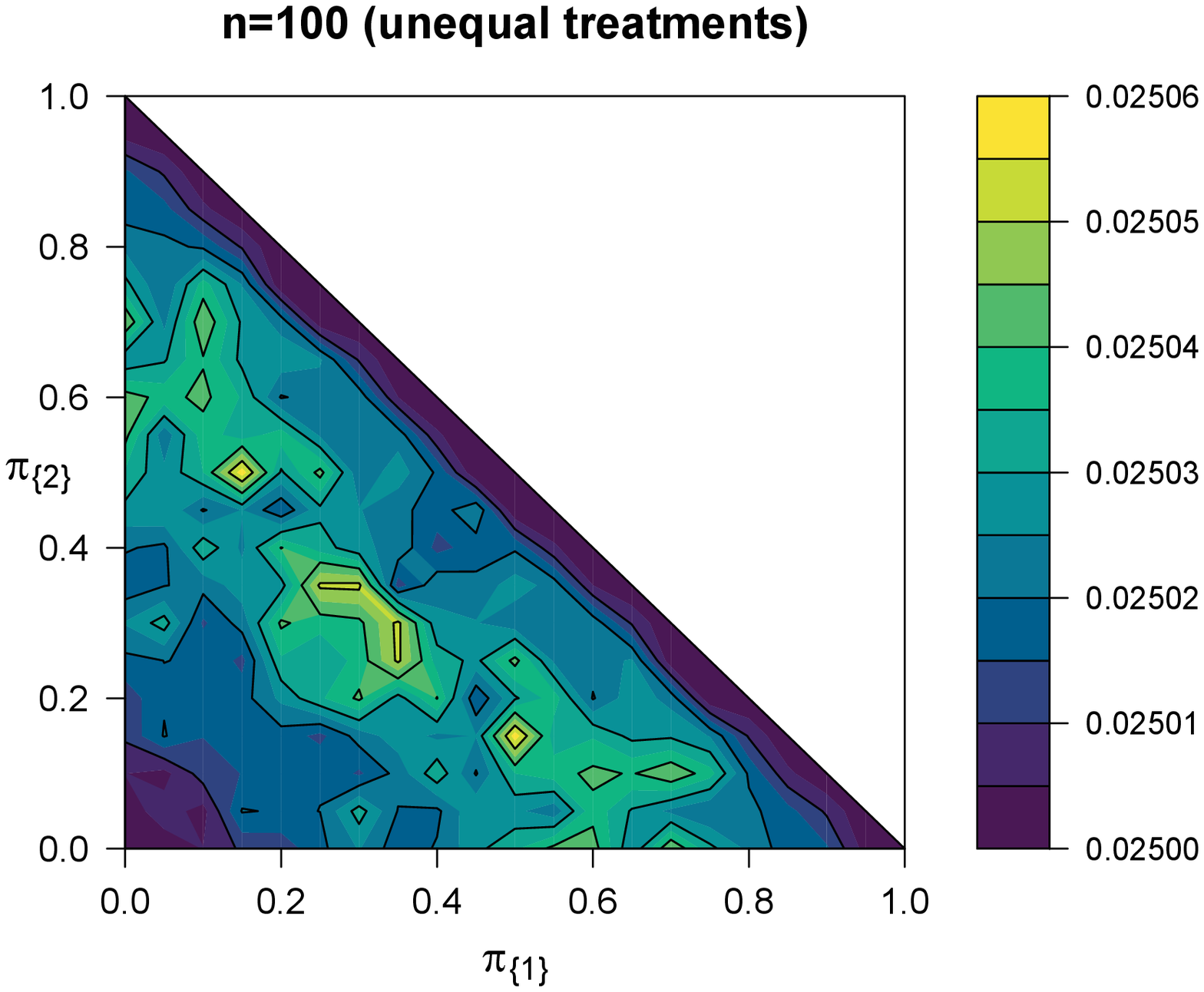} 
  \end{minipage} \hfill
\caption{\label{fig: contourplots} Contour plots of the actual overall PWER when using ML-estimates $\hat{\pi}_J$ for the prevalences $\pi_J$ in the determination of the critical value $c^\ast_\PWERc$ at level $\alpha=0.025$. The first row corresponds to scenario (i) and the second row to scenario (ii) from Section 4.2. 
Because of $\pi_{\{1\}}+\pi_{\{2\}}\le 1$,  the contour plots are restricted to the lower left rectangle of the squares.} 
\end{figure}

We examine the PWER by means of scenarios (i) and (ii) of Section~\ref{sec_OneStudy}. For each constellation 
of true prevalences, we generate sample size vectors $(\hat{n}_J)_{J\subseteq I}$ from the corresponding multinomial distribution and computed the MLEs $(\hat{\pi}_J)_{J\subseteq I}$. 
To see by how much the true PWER is inflated, the probabilities for a type I error for each sub-population $\Pop_J$ are computed by using the ``estimated'' critical value and the conditional correlation structure of the involved test statistics. By weighting each of these probabilities by their respective true population prevalence $\pi_J$, we obtain 
the true PWER for the given ``estimated'' critical value. This procedure was repeated 10.000 times and the mean of each true PWER was taken as approximation of the actual overall PWER.  Fig.~\ref{fig: contourplots} shows contour plots of this approximation of the overall PWER for scenarios (i) and (ii) and $N=50$ and $N=100$, respectively. The plots indicate that the target PWER of $0.025$ may be missed only slightly, even for $N=50$. 

In case of a small relative population size $\pi_J$ it might be that, by chance, no patient is recruited in this group. This would mean that we would not account for all the multiplicity for these patients completely. If such intersection cannot be excluded theoretically from the inclusion and exclusion criteria or due to medical arguments, we could introduce a small minimal number $\pi_{min}$ for all $\pi_J$ in order to be more conservative. Also different approaches like shrinkage methods or Bayesian estimation of the $\pi_J$ are conceivable options in future research.

\subsection{Multiple testing approaches for umbrella trials}\label{sec: mtfut}

We consider now a multiple testing approach for umbrella trials suggested in Sun \textit{et al.} (2016) and investigate
the gain in power by switching from FWER- to PWER-control. Following Sun \textit{et al.} (2016), we assume $l$ disjoint population strata, which are denoted here by 
$\Str_1,\dots,\Str_l$. In each stratum a specific experimental treatment $E_i$ \textcolor{black}{shall be} compared to a common control $C$. For simplicity, we assume that each population has the \textcolor{black}{relative population prevalence $\pi_i$ which is assumed to equal $n_i/N$}, where $n_i$ is the number of patients in $\Str_i$ and $N$ is the total number of patients. This holds in practice at least approximately; see also Section~\ref{sec: eopp}. 

With only small $n_i$, the establishment of a treatment effect in the individual strata is difficult and impossible to achieve with sufficient power. Therefore, study designs have been suggested that compare the global treatment strategy $E$ which assigns treatment $E_i$ to population strata $\Str_i$, as a total with the control treatment in the overall population. Such an overall comparison of the strategy $E$ with $C$ utilizes the total sample size $N$ and does also not require multiple testing. However, it does not permit a claim for a sub-population when the effect of $E$ is heterogeneous. To improve the approach, Sun \textit{et al.} (2016) suggest to test all sub-strategies 
$E^S$, $S\subseteq \{1,\ldots,l\}$, that consider only the union $\Pop^S=\cup_{i\in S}\Str_i$ with treatment assignments as in $E$, against the control in 
$\Pop^S$. This permits claims also for sub-populations and thereby increases the possibility for
the efficacy conclusions. 
Of course, such testing requires an adjustment for multiplicity. Sun \textit{et al.} (2016) provide a (single-step) procedure that controls the FWER.

For the formal description of the procedures, let $\boldsymbol\theta = (\theta_1, \dots, \theta_l)$ be the vector of unknown treatment effects (mean differences) in the populations, and consider for each $S\subseteq \{1,\ldots,l\}$ the average treatment effect in $\Pop^S$:
$$\theta^S = \sum_{i\in S}(\pi_i/\pi^S)\theta_i$$ 
with $\pi^S = \sum_{i\in S}\pi_i$ the relative prevalence of $\Pop^S$. Sun \textit{et al.} (2016) assume the linear model 
\begin{align}\label{linmod}
Y_{ij} = \mu_i + \theta_i X_{ij} + \varepsilon_{ij},
\end{align}
where $X_{ij}$ denotes the treatment indicator for patient $j$ in group $i$ which equals 1 if assigned to the experimental treatment $E_i$ and otherwise $0$, and $\theta_i$ is the treatment effect of $E_i$ in population $\Str_i$. The error terms $\varepsilon_{ij}$ are assumed to be i.i.d.\ normally distributed with mean $0$ and homogeneous variance $\sigma^2$. As mentioned above, the authors suggest to test
\begin{align}
H^S: \theta^S \le 0 \quad \mbox{vs.} \quad K^S: \theta^S > 0 \quad \text{for all } S \subseteq L=\{1,\ldots,l\}. 
\end{align}
Note that the $\Pop^S$ and $H^S$, $S\subseteq L$, correspond to the $\Pop_i$ and $H_i$, $i\in I$, in Section~\ref{sec PWER} and \ref{sec_CofPWER}.   

From the least squares estimate of the linear model, we obtain one-sided t-test statistics $T^S$ for testing $H^S$ for each $S\subseteq L$. In order to control the FWER, \textcolor{black}{Sun \textit{et al.} (2016)} conduct a single-step procedure that compares each $T^S$ with the upper $\alpha$-quantile $c^\ast_\FWERc$ of the distribution of $\max{\{T^S\,\vert\, S\subseteq L\}}$ under the global null hypothesis, i.e.\ the assumption that none of the treatments $E_i$ is superior to the control. We finally select the subset $S^*_\FWERc\subseteq L$ for which a positive treatment effect is claimed and that yields the largest value of $T^S$,
\begin{align}
S^*_\FWERc = 
\begin{cases} 
\argmax_{S\subseteq I} T^S, & \text{if $\max{\{T^S\,\vert\, S\subseteq L\}} > c^*_\FWERc$}\\
\emptyset, & \text{else.}
\end{cases}
\end{align}

To achieve PWER-control at the same level $\alpha$, we determine the critical value $c_{\text{\PWERc}}^*$ such that $\PWER= \alpha$ holds under the global null hypotheses. While $\Str_1,\dots, \Str_l$ are disjoint, some of their unions $\Pop^S$overlap. Since not all $\Pop^S$ overlap, the FWER corrects the multiple type I error rate for cases that cannot occur (similar to example 3) and hence may be viewed as overly conservative.

The PWER under the global null hypothesis ($\boldsymbol\theta = \textbf{0}=(0,\ldots,0)$) is given by
\begin{align}
\label{eq: PWERcap4}
\PWER_{\textbf{0}} &= \sum_{i=1}^l\pi_i\mathbb{P}_{\textbf{0}}\left(\bigcup_{S\ni i}\left\{T^S \ge c_{\PWERc}^*\right\}\right),
\end{align}
where ``$S\ni i$'' denotes all $S\subseteq L$ that contain the index $i$. This is because population $\Str_i$  is affected by a type I error whenever a hypothesis $H^S$  is erroneously rejected that corresponds to a population $\Pop^S$ for which $i\in S$ (or $\Str_i \subseteq \Pop^S$).  

\nocite{bretz2016multiple,mvtnorm}

Due to the assumption of a homogeneous residual variance and the $2l$ mean parameter in the linear model (\ref{linmod}), $\{T^S\}_{S\subseteq L}$ follows a joint t-distribution with $N-2l$ degrees of freedom. 
In R, the distribution function of the multivariate t-distribution is implemented in the \texttt{mvtnorm}-package (see Genz \textit{et al.}, 2017) via the function \texttt{pmvt} which needs the degrees of freedom  and the correlation matrix of the test statistics as input (see e.g.\ Bretz \textit{et al.}, 2016). The correlation matrix 
can be computed using the contrast matrix and the design matrix of the linear model. Probabilities in (\ref{eq: PWERcap4}) are then calculated by choosing the appropriate sub-matrices of the correlation matrix. Thus, for known values of $\pi_i$, $i\in L$, and $l$, we can numerically determine the critical value $c_{\PWERc}^*$ such that $\PWER = \alpha$.

We know that $c_{\FWERc}^* > c_{\PWERc}^*$, which implies that whenever the FWER-approach selects a non-empty $S^*_\FWERc$, the same set is selected by the PWER-approach, $S^*_\PWERc=S^*_\FWERc$. We may, however, select the empty set with the FWER-approach, $S^*_\FWERc=\emptyset$, while $S^*_\PWERc\not=\emptyset$.\\

\textbf{Performance measures:}
Sun \textit{et al.} (2016) examined several quality and performance measures to assess how good a selected subset $S^*$ is. 
For example, they considered the average effect in the overall population when applying treatment strategy $E^{S^\ast}$ in $\Pop^{S^\ast}$ and the control in the rest of the population. 
We will consider the relative quantity $\text{RAE}= 100\,\mathbb{E}\left(\sum_{i\in S^*}\pi_i\theta_i\right)/\theta_\text{overall}$ where $\mathbb{E}$ is the expectation with respect to the sample
distribution \textcolor{black}{and $\theta_\text{overall}$ is the weighted average of the positive treatment effects, 
$$\theta_\text{overall} = \sum_{i\in L_+}\pi_i\theta_i/\sum_{i\in L_+} \pi_i \quad\text{for}\quad L_+=\{i=1,\ldots,l:\theta_i>0\},$$
that describes how efficient the experimental treatment strategy $E$ is for the union of sub-populations that benefit from $E$.} Since the PWER-procedure chooses a non-empty $S^*$ more often as the FWER-procedure, this quantity will always be larger for the PWER-approach. 

In addition to this measure we will investigate the average size of the `correctly' chosen subgroups within the selected ones, i.e.\ the average of $\pi^{S^*_+}/\pi^{S^*}$ where $S^\ast_+={\{i\in S^*| \theta_i > 0\}}$
and $\pi^{S^\ast_+}=\sum_{i\in S^\ast_+} \pi_i$. This gives the fraction of the patient cohort that benefits from the experimental treatment strategy within the one that is exposed to $E^{S^\ast}$ by the results of the study.
Analogously, we are interested in the average of the relative size of the `falsely' chosen subgroups within the chosen ones: $\pi^{S^*_0}/\pi^{S^*}$ with 
$S^\ast_0={\{i\in S^*| \theta_i \le 0\}}$.
Lastly, we consider the probability of rejecting at least one false null hypothesis, 
$$\text{Power}=\mathbb{P}(\text{\,reject any $H^S$ with $\theta^S > 0$, $S\subseteq L$}\,),$$ 
as a way to measure the power of the procedures.\\ 


\textbf{Design of the simulation:} To make our results comparable to those of Sun \textit{et al.} (2016), we conducted simulations with roughly the same parameters. That is, for the cases of $l = 2,4,6$ sub-populations and a significance level $\alpha = 0.025$, we chose a total sample size of $N = 1056$ and assume that all group-specific intercepts $\mu_i$ are equal to 0. Also, for simplicity, each group is assumed to be of equal size, i.e.\  $\pi_1 = \dots = \pi_l$. 

As in Sun \textit{et al.} (2016), we assume non-negative effects $\theta_i\ge 0$ and choose $\boldsymbol\theta = (\theta_1,\dots, \theta_l)$ based on the number of subgroups $l$ and three further characteristics. The first one is the percentage of true null hypothesis: $q=l_0/l$ with $l_0$ the size of $L_0=\{i=1,\ldots,l:\theta_i= 0\}$. The second one characterizes the treatment effect heterogeneity and is defined as
$$\tau=(\theta_\text{max}-\theta_\text{min})/(\theta_\text{max}+\theta_\text{min})$$ 
where $\theta_\text{max}=\max_{i\in L_+}\theta_i$ and $\theta_\text{min}=\min_{i \in L_+}\theta_i$. Note that $\tau$ 
equals the relative half-range of the positive $\theta_i$'s, i.e.\ half of their range divided by the average of their extremes. Obviously, a large $\tau$ means a large heterogeneity between the positive $\theta_i$. \textcolor{black}{The third one is the weighted average $\theta_\text{overall}$ as previously introduced.}

Given values for $l$, $q$, $\theta_\text{overall}$ and $\tau$ one finds a gird of $l$ equidistant points such that the three characteristics are met. One easily verifies, that this grid is uniquely determined by the four quantities. 
\textcolor{black}{Following Sun \textit{et al.}} we chose $q$ such that $q\cdot l$ is always an integer. \textcolor{black}{Note that for $q\ge (l-1)/l$ there is at most one $\theta_i\not=0$ and so $\tau=0$ 
(no heterogeneity) is the only possible value for $\tau$.}\\


\textbf{Results:} The simulation results for $l=2$ and $4$ are given in Table~\ref{tab: tab1} and for $l=6$ and $8$ in Appendix C. On can see from the tables that control of the PWER, in comparison to FWER-control, provides a substantially  
larger power and larger average proportion of `correctly' chosen subgroups and a larger average effect. It also increases the proportion of `falsely' chosen subgroups. This is because a subgroup is selected more frequently with PWER-control. 

While the proportion of `falsely' chosen subgroups is increased by at most 2.2\% (percentage points) and remains below 5\% (one-sided), the proportion of `correctly' chosen subgroups (among the selected ones) and the power are increased by up to 10\%   
and often by more than 5\%. The expected effect RAE is always larger with PWER-control. 

Under the global null hypothesis ($P=1$) the
average proportion of `falsely' selected populations equals by theory the one-sided family-wise error rate. With PWER-control at level 2.5\% the FWER was found to be between 3.6\% and 4.5\% for $l=2,4,6,8$. Note that the average proportion of `falsely' selected populations exceeds the level of 2.5\% (sometimes substantially) also with FWER-control when there is an effect in some but not all population strata.
 
In summary, we see that control of PWER substantially increases the chance for a delivery of efficient treatments while the risk of receiving an inefficient treatment and the percentage of patients that do not benefit from the treatment decisions is increased to a moderate extend and remains comparable to the procedure with FWER-control.      

\section{Extension to simultaneous confidence intervals}\label{sec: SCI}

We are coming back to the general set-up of Section~\ref{sec PWER} and \ref{sec_CofPWER}.  Utilizing the duality between (multiple) hypothesis tests and (simultaneous) confidence intervals, 
the multiple test procedure with control of the PWER, introduced in Section~\ref{sec_CofPWER}, can be extended to confidence intervals for the efficacy parameter  
$\theta_i=\theta(\Pop_i,T_i)$, $i=1,\ldots, m$.  In this section we will introduce the dual simultaneous 	confidence intervals and discuss their coverage properties.

To introduce the confidence intervals, let $\boldsymbol\delta=(\delta_1,\ldots,\delta_m)$ be a vector of possible values for $\boldsymbol\theta=(\theta_1,\ldots,\theta_m)$ and consider the corresponding null hypotheses $\Hd{i}: \theta_i= \delta_i$, $i=1,\ldots,m$. Assume further that $\Td{i}$, $i=1,\ldots,m$, are (asymptotically) pivotal test statistics for $\Hd{i}$, i.e., 
the (asymptotic) joint distribution of $(\Td{1},\ldots,\Td{m})$ under $\boldsymbol\theta=\boldsymbol\delta$ is the same for all $\delta$. If $\Td{i}$ decreases in $\delta_i$ for the given data, 
then it makes sense to form the one-sided intervals $\mC_i=[\ttheta_i,\infty[$ with the lower bound
\begin{align}
\label{eq ttheta}
\ttheta_i:=\min \{\delta_i: \Td{i}\le c^\ast\}
\end{align} 
where $c^\ast$ is the critical value defined in (\ref{eq: PWEReq}) for $\boldsymbol\theta^\ast=\boldsymbol\delta$. Because $(\Td{1},\ldots,\Td{m})$ is pivotal, the critical value $c^\ast$ is independent from $\boldsymbol\delta$. 
The monotonicity of $\Td{i}$ applies to most (one-sided) tests and is satisfied e.g.\ for Wald-type test statistics $\Td{i}=(\htheta_i-\delta_i)/SE_i$ where $\hat{\theta}_i$ is an estimate of $\theta_i$ (e.g.\ the MLE) with an standard error $SE_i$ that is independent of the parameter value $\delta$. In this case we obtain
$\ttheta_i=\htheta_i-c^\ast SE_i$.

Upper confidence bounds can be derived by applying the same principle and two-sided confidence intervals are obtained by the intersection of the two one-sided intervals. With Wald-type dual tests we obtain the two-sided intervals $\mC_i=\left[\htheta_i-c^\ast SE_i\,,\,\htheta_i+c^\ast SE_i\right]$. 

We finally discuss the coverage properties of the above introduced confidence bounds and intervals. We start with the lower confidence bounds $\ttheta_i$. To this end, consider a patient $P$ that is randomly drawn from $\Pop$ and let $I_P$ be the set of indices of the sub-populations $\Pop_i$ the patient $P$ belongs to, i.e.\ $I_P=\{i: P\in\Pop_i\}$. The set $I_P$ gives all population efficacy parameter $\theta_i$, $i\in I_P$, that are relevant for patient $P$. Note that $I_P$ is a random set, because $P$ is randomly drawn from 
$\Pop$. If $\theta_i$ is the true unknown efficacy parameter, then by the definition (\ref{eq ttheta}) we get $\ttheta_i>\theta_i$ if and only if $T^{\theta_i}_{i}> c^\ast$.  
Since the dual tests for $H^{\theta_1}_1,\ldots,H^{\theta_m}_m$ control the PWER, the (simultaneous) probability that any of the lower confidence bounds $\ttheta_j$, $j\in I_P$, fall above the true $\theta_j$ is at most $\alpha$. This gives the coverage property
\begin{align}\label{eq copb}
\mathbb{P}_\theta\left( \ttheta_j \le \theta_j \mbox{ for all }  j\in I_P\right) \ge 1-\alpha\, 
\end{align}
meaning that with a probability of at most $1-\alpha$, for a randomly chosen patient, the lower confidence intervals $[\ttheta_j,\infty[$, $j=1,\ldots, m$, cover all true $\theta_j=\theta(\Pop_j,T_j)$ that are relevant to this patient. 
Because, $I_P=J$ if and only if $P\in\Pop_J=\cap_{j\in J} \Pop_j\setminus \bigcup_{k\in I\setminus J}\Pop_k$ we can write the coverage probability as
$$\sum_{J\subseteq I} \pi_J \mathbb{P}_\theta\left( \ttheta_j \le \theta_j \mbox{ for all }  j\in J\right).$$
Hence, equation  (\ref{eq copb}) means to control a kind of average simultaneous coverage probability where we focus in each stratum on the relevant confidence statements and average the strata-wise coverage probability over the entire population $\Pop$.  

The upper confidence bounds and two-sided confidence intervals control the same type of average simultaneous coverage probability. As for the classical confidence intervals, the two-sided interval have a twice as large non-coverage probability as the one-sided intervals.

\begin{table}
\caption{\label{tab: tab1} Simulation results for $l=2$ and $l=4$. Results for power (\%), the percentage of correctly and falsely chosen sub-populations and the relative average effect (RAE) for PWER- and FWER-control under parameter configurations $\boldsymbol\theta = (\theta_1,\dots, \theta_l)$ that depend on the fraction of true null hypotheses $q$ and the relative half-range $\tau$ of the positive $\theta_i$`s. }
\centering
\begin{tabular}{rrrrrrrrrrrrr}
\toprule
& && Power & correct & false & RAE && Power & correct & false & RAE\\
\cmidrule{1-12}
&$l=2$ && \multicolumn{4}{l}{$q=0$} && &&&\\
\cmidrule{1-7}
\multirow{2}{*}{$\tau=0$} & PWER && 36.4 & 36.4 & 0 & 2.9 && & & & \\
& FWER && 31.0 &31.0 &0 &2.5 & & & &\\
\cmidrule{1-7}
\multirow{2}{*}{$\tau=0.4$} & PWER && 40.4 &40.4 &0 &3.3 && & & & \\
& FWER && 34.6 &34.6 &0 &2.8 && & & &\\
\cmidrule{1-7}
\multirow{2}{*}{$\tau=0.8$} & PWER &&51.2 &51.2 & 0& 4.7&& & & & \\
& FWER && 45.2& 45.2& 0& 4.2&& & & &\\
\cmidrule{1-12}
& $l=2$ && \multicolumn{4}{l}{$q=1/2$} && \multicolumn{4}{l}{$q=1$}\\
\cmidrule{1-12}
\multirow{2}{*}{$\tau=0$} & PWER && 57.7 & 52.9 &4.8 &5.8 &&0 &0 &3.6 &0 \\
& FWER &&52.0 &47.8 &4.2 &5.2 &&0 &0 &2.4 &0\\
\cmidrule{1-12}
\\[-1.5em]
\cmidrule{1-12}
& $l=4$ && \multicolumn{4}{l}{$q=0$} && \multicolumn{4}{l}{$q=1/4$}\\
\cmidrule{1-12}
\multirow{2}{*}{$\tau=0$} & PWER &&36.2 &36.2 &0 &2.3 &&42.2 &38.8 &3.5 &3.0 \\
& FWER &&27.4 &27.4 &0 &1.7 &&32.7 &30.1 &2.6 &2.4\\
\cmidrule{1-12}
\multirow{2}{*}{$\tau=0.4$} & PWER &&37.9 &37.9 &0 &2.5 &&44.8 &41.3 &3.6 &3.3 \\
& FWER &&29.1 &29.1 &0 &1.9 &&35.5 &32.7 &2.7 &2.6\\
\cmidrule{1-12}
\multirow{2}{*}{$\tau=0.8$} & PWER && 43.0&43.0 &0 &3.0 &&52.7 &48.9 &3.7 &4.2 \\
& FWER && 33.7& 33.7& 0& 2.4 && 43.2 &40.2 &3.0 &3.5\\
\cmidrule{1-12}
& $l=4$ && \multicolumn{4}{l}{$q=2/4$} && \multicolumn{4}{c}{}\\
\cmidrule{1-7}
\multirow{2}{*}{$\tau=0$} & PWER &&53.2 &45.7 &7.6 &4.6 &&&&& \\
& FWER &&43.8 &37.8 &6.1 &3.8 &&&&& \\
\cmidrule{1-7}
\multirow{2}{*}{$\tau=0.4$} & PWER &&58.8 &51.1 &7.8 &5.0 && & & & \\
& FWER &&49.5 &43.1 &6.4 &4.2 && & & &\\
\cmidrule{1-7}
\multirow{2}{*}{$\tau=0.8$} & PWER && 73.9&65.9 &8.0 &6.8 && & & & \\
& FWER &&65.3 &58.5 &6.8 &6.0 && & & &\\
\cmidrule{1-12}
& $l=4$ && \multicolumn{4}{l}{$q=3/4$} && \multicolumn{4}{l}{$q=1$} \\
\cmidrule{1-12}
\multirow{2}{*}{$\tau=0$} & PWER &&81.5 &70.1 &11.5 &8.1 &&0 &0 &4.2 &0 \\
& FWER &&75.1 &64.9 &10.2 &7.5 &&0 &0 &2.4 &0 \\
\cmidrule{1-12}
\end{tabular}
\end{table}

\section{Discussion}\label{sec: disc}

This paper introduces a new multiple type I error rate concept for clinical trials with multiple and possibly intersecting populations that permits for more liberal and more powerful tests than control of the family-wise error rate (FWER). It relies on the observation that not all patients are affected by all test decisions, since not all hypotheses concern all population strata. By averaging the individually relevant, multiple type I errors over the entire population, it provides control of the probability that a randomly selected patient will be exposed to an inefficient treatment strategy. This average multiple type I error rate, which we call the {\em population-wise error rate (PWER)}, is smaller than the maximum multiple type I error rate a patient is exposed to.

Let us recall that we only consider population-wise claims, i.e.\ claims on {\em treatment strategies} that consist of a treatment and a population the treatment is intended for and for which the average treatment effect is the estimand of interest. This is also the case when going for FWER control. No individual efficacy claims are anticipated here. Error control of patient-wise claims is impossible without sacrificing power or making strong assumptions. However, a population-wise claim can be viewed as a proxy or approximation for individual claims in the target population. Test results from more than a single population may be used for a more informed individual decision. With PWER control, we consider the worst case scenario, where an efficacy claim for a treatment strategy will always lead to an application of the treatment to all patients in the target population. 
Note that we do not account for a potential off-label use where a treatment is applied to patients outside its target population. 

\nocite{Liu2014CommentAF} 

We have presented a simple approach for achieving PWER-control by adjustment of critical values and have illustrated the power gain when passing from FWER to PWER in a number of examples. We have mainly considered the simple situation of multivariate normal distributed test statistics. This situation applies at least asymptotically to a large number of hypothesis tests for which PWER control is then guaranteed asymptotically. The methods and principle introduced here can also be implemented with other finite sample distributions like e.g. the multivariate t-distribution (as done in Section \ref{sec: mtfut}) or be improved via resampling methods. Variance heterogeneity across populations is a general issue for trials with multiple populations that applies similarly to procedures with FWER control (see e.g.\ Placzek and Friede, 2019). One can say, whenever control of the FWER is possible then control of the PWER is possible as well, since the latter just controls an average of family-wise error rates.
We have also extended the suggested multiple test to simultaneous confidence intervals and showed that these intervals control, for a randomly chosen patient, the probability of a simultaneously correct statement on the parameters that are relevant for this individual. 

Control of the PWER requires the knowledge of the relative prevalences of all disjoint population strata. These may either be obtained from previous studies or may be estimated at the end of the study. This complicates PWER control. We have illustrated in an example with two populations that the estimation of the prevalences does not strongly harm PWER control even with moderate sample sizes. However, more examples with more hypotheses are required to fully explore this issue.  At least, PWER control is always guaranteed asymptotically.

Since our procedure simply results in an adjustment of critical values, power calculations and power simulations are straightforward and deviate only minimally from approaches for classical multiple tests, except for the fact that the critical values may depend on the sample via the prevalence estimates. This can be resolved by using a priori estimates of the prevalences based on experience and past studies. The same issues arises from the estimation of the correlation structure of the test statics used for an efficient PWER and FWER control. A miss-specification of the prevalences may  be corrected in a mid-trial blinded sample size review (Placzek and Friede, 2018). 

In Section~\ref{sec_CofPWER} we have suggested a single-step procedure to control the PWER and one might ask whether this procedure can be uniformly improved by a step-down test because this is the case for single-step tests with FWER control (e.g.\ Dmitrienko \textit{et al.}, 2009). For instance, in Example~\ref{sec ex1} with two intersecting hypotheses, we may ask whether we can test $H_2$ with a smaller critical $c^\ast_2<c^\ast$ when $H_1$ has already been rejected with critical value $c^\ast$.
One can quickly see that this is not possible. To this end assume that both hypotheses $H_1$ and $H_2$ are true. Rejection of $H_2$ when $Z_2\ge c^\ast$ or $Z_2\ge c^\ast_2$ with $Z_1\ge c^\ast$, obviously increases the second and third terms in equation (\ref{eq: ex1PWERboth}) of the PWER. Since we have chosen $c^\ast$ to be the smallest critical value that satisfies (\ref{eq: PWEReq}), which leads to an PWER equal to $\alpha$ with continuously distributed $Z_i$ (a generic and common situation), we do not control the PWER for any $c^\ast_2<c^\ast$. 
We may define PWER-controlling step-down tests with an enlarged $c^\ast$ in order to mimic and improve step-down tests with FWER control. However, such procedures do not uniformly improve the single-step test with PWER control and are therefore beyond the scope of this paper. The development of step-down tests with PWER control is a topic of future research.     

\nocite{dmitrienko}

Single-step procedures have the advantage that they can directly be extended by simple and well behaving simultaneous confidence intervals (SCIs). We have illustrated this in Section~\ref{sec: SCI} for single-step tests with PWER control. An extension to simple and well behaving SCIs is impossible for step-down tests: Compatible SCIs often are non-informative in the sense that they do not provide any additional information to the sheer hypothesis tests (Strassburger and Bretz, 2008; Guildbaud, 2009) and sufficiently informative SCIs are compatible only to a modification of the original step-down test (Brannath and Schmidt, 2014). This justifies the use of single step tests in practice.     

\nocite{StrassbugerBretz,Guilbaud,BrannathSchmidt}

We finally remark that an extension of the presented PWER approach to multi-stage and adaptive designs is under development by the authors and will be a topic of future contributions. Multi-stage and particularly flexible designs 
provide the opportunity for adding or dropping populations at interim analyses based on the unblinded interim data (e.g.\ Brannath \textit{et al.}, 2009; Wassmer and Brannath, 2016; Placzek and Friede, 2019). In the example of Section~\ref{sec ex1} we may for instance add and enrich the intersection of the two populations for an investigation in a second stage of the study if the efficacy of the treatment is seen at interim only in one of the two populations. Hence, the development of adaptive and sequential designs with PWER control is an interesting and valuable research task.

\nocite{BrannathEtAl,WassmerBrannath,PlaczekFriede1,PlaczekFriede2}

\section*{Acknowledgements}
The authors want to thank Dr.\ Miriam Kesselmeier at the University Hospital of Jena for her constructive comments on a previous version of this manuscript.
\section*{Funding}
This research was supported by the BMBF under the funding number 01EK1503B. 
\appendix

\section{Derivation of $c^*_{\PWERc}$ in Section 4.1}
The solutions of the quadratic equation $1- (1-\pi_{\{1,2\}})\Phi(c_{\PWERc}^*)- \pi_{\{1,2\}} \Phi(c_{\PWERc}^*)^2 = \alpha $
are:
\begin{align*}
x_{1/2} = \frac{-(1-\pi_{\{1,2\}})\mp \sqrt{(1-\pi_{\{1,2\}})^2+4\pi_{\{1,2\}}(1-\alpha)}}{2\pi_{\{1,2\}}}.
\end{align*}
Since $\sqrt{(1-\pi_{\{1,2\}})^2+4\pi_{\{1,2\}}(1-\alpha)} > (1-\pi_{\{1,2\}})$ for all $\pi_{\{1,2\}}\in(0,1]$ it follows that 
$$c^*_{\PWERc} = \Phi^{-1}(x_2) = \Phi^{-1}\left(\frac{-(1-\pi_{\{1,2\}})+ \sqrt{(1-\pi_{\{1,2\}})^2+4\pi_{\{1,2\}}(1-\alpha)}}{2\pi_{\{1,2\}}}\right)$$ is the only valid solution. 
To show that $c^*_{\PWERc}$ is strictly monotonically increasing in $\pi_{\{1,2\}}$, we consider the function   
$y=y(\pi_{\{1,2\}})=\Phi(c^*_{\PWERc})\in(0,1)$ which satisfies the equation 
\begin{align*}
\alpha = 1-y+\pi_{\{1,2\}}y - \pi_{\{1,2\}}y^2.
\end{align*}
Taking derivatives w.r.t.\ $\pi_{\{1,2\}}$ on both sides of this equation yields after a rearrangement of terms:
\begin{align*}
y' &= y(1-y)/\{1+\pi_{\{1,2\}}(2y-1)\}.
\end{align*}
Due to $\pi_{\{1,2\}}(2y-1) \ge -1$ it follows that $y' > 0$ for all $\pi_{\{1,2\}}\in(0,1)$. 
\section{Calculation of the correlation expressions in Section 4.2}
The variance of $\hat{x}_{T_i} - \hat{x}_{C_i}$ is easily found by exploiting the independence of the individual observations and is given by 
$\text{Var}(\hat{x}_{T_i}-\hat{x}_{C_i}) =  (2\sigma^2/N) v^2_i$ with
\begin{align*}
v^2_i= \left({\pi_{\{i\}}}/{\pi_i}\right)^2\left({2}/{\pi_{\{i\}}}\right) +  
\left({\pi_{\{1,2\}}}/{\pi_{i}}\right)^2\left({3}/{\pi_{\{1,2\}}}\right)\quad \text{where}\quad \pi_i = \pi_{\{i\}} + \pi_{\{1,2\}}.
\end{align*}

We turn to the correlation between $Z_1$ and $Z_2$ for the case (i) of two different treatments $T_1  \not= T_2$.
By the independence of means from disjoint cohorts, we calculate
\begin{align*}
\text{Cov}(Z_1,Z_2) &= \frac{\text{Cov}(\hat{x}_{T_1}-\hat{x}_{C_1}, \hat{x}_{T_2}-\hat{x}_{C_2})}{2\sigma^2 v_1 v_2/N} 
= \frac{\text{Cov}(\hat{x}_{C_1}, \hat{x}_{C_2})}{2\sigma^2 v_1 v_2/N}\\ 
&= \frac{\text{Cov}\left(\frac{\pi_{\{1\}}}{\pi_1}\bar{x}_{C,\{1\}}+ \frac{\pi_{\{1,2\}}}{\pi_1}\bar{x}_{C,\{1,2\}}, \frac{\pi_{\{2\}}}{\pi_2}\bar{x}_{C,\{2\}}+ \frac{\pi_{\{1,2\}}}{\pi_2}\bar{x}_{C,\{1,2\}}\right)}{2\sigma^2 v_1 v_2/N}\\
&= \frac{\pi_{\{1,2\}}^2}{\pi_1\pi_2}\frac{\text{Cov}(\bar{x}_{C,\{1,2\}},\bar{x}_{C,\{1,2\}})}{2\sigma^2 v_1 v_2/N} = \frac{\pi_{\{1,2\}}^2}{\pi_1\pi_2}\frac{3\sigma^2/n_{\{1,2\}}}{2\sigma^2 v_1 v_2/N} = \frac{3\pi_{\{1,2\}}}{2\pi_1\pi_2 v_1 v_2}
\end{align*}
where we used $n_J = N\pi_J$ for $J\subseteq\{1,2\}$ in the last equation. Now, if $\pi_{\{1\}} = \pi_{\{2\}}$ then $\pi_{\{1\}} = \pi_{\{2\}} = (1-\pi_{\{1,2\}})/2$ and $\pi_1=\pi_2 = (1+\pi_{\{1,2\}})/2$, and the correlation reduces to
\begin{align*}
\text{Cov}(Z_1,Z_2) &= \frac{6\pi_{\{1,2\}}}{(1+\pi_{\{1,2\}})^2\left\{\left(\frac{1-\pi_{\{1,2\}}}{1+\pi_{\{1,2\}}}\right)^2\left(\frac{4}{1-\pi_{\{1,2\}}}\right)+ \left(\frac{2\pi_{\{1,2\}}}{1+\pi_{\{1,2\}}}\right)^2\left(\frac{3}{\pi_{\{1,2\}}}\right)\right\}}\\
&= \frac{6\pi_{\{1,2\}}}{4(1-\pi_{\{1,2\}})+12\pi_{\{1,2\}}}= \frac{3\pi_{\{1,2\}}}{2(1+2\pi_{\{1,2\}})}.
\end{align*}
For case (ii), where $T_1=T_2 = T$ we calculate
\begin{align*}
\text{Cov}(Z_1,Z_2) &= \text{Cov}(\bar{x}_{T,1}-\bar{x}_{C,1}, \bar{x}_{T,2}-\bar{x}_{C,2})\frac{\sqrt{n_1n_2}}{4\sigma^2}
= \text{Var}(\bar{x}_{T,\{1,2\}}-\bar{x}_{C,\{1,2\}})\frac{n^2_{\{1,2\}}}{4\sigma^2\sqrt{n_1 n_2}}\\
&= {n_{\{1,2\}}}/{\sqrt{n_1n_2}} = {\pi_{\{1,2\}}}/{\sqrt{\pi_1\pi_2}},
\end{align*}
and for $\pi_{\{1\}} = \pi_{\{2\}}$ we obtain $\text{Corr}(Z_1, Z_2) = {2\pi_{\{1,2\}}}/{(1+\pi_{\{1,2\}})}$.
Obviously, this correlation is greater than the one from case (i) for all $\pi_{\{1,2\}}\in[0,1]$. 

\newpage
\section{Further simulation results}
\begin{tabular}{l l l l l l l l l l}
\toprule
$l=6$ & & Power & correct & false & RAE & Power & correct & false & RAE\\
\cmidrule{1-10}
& & \multicolumn{4}{l}{$q=0$} & \multicolumn{4}{l}{$q=1/6$}\\
\cmidrule{1-10}
\multirow{2}{*}{$\tau=0$} & PWER & 34.9&34.9 &0 &2.0 &37.4 &34.9 &2.5 &23 \\
& FWER &26.0 &26.0 &0 &1.5 &28.1 &26.3 &1.8 &17\\
\cmidrule{1-10}
\multirow{2}{*}{$\tau=0.4$} & PWER &36.3 &36.3 &0 &2.1 &39.1 &36.4 &2.7 &24 \\
& FWER &26.6 &26.6 &0 &1.6 &29.2 &27.3 &1.9 &19\\
\cmidrule{1-10}
\multirow{2}{*}{$\tau=0.8$} & PWER &39.9 &39.9 &0 &2.5 &43.6 &40.9 &2.8 &30 \\
& FWER &29.9 &29.9 &0 &1.9 &33.6 &31.5 &2.1 &23\\
\cmidrule{1-10}
& & \multicolumn{4}{l}{$q=2/6$} & \multicolumn{4}{l}{$q=3/6$}\\
\cmidrule{1-10}
\multirow{2}{*}{$\tau=0$} & PWER &42.0 &36.6 &5.4 &2.8 &49.8 &40.9 &9.0 &38 \\
& FWER &32.3 &28.2 &4.1 &2.2 &39.5 &32.6 &7.0 &31\\
\cmidrule{1-10}
\multirow{2}{*}{$\tau=0.4$} & PWER &44.1 &38.5 &5.6 &3.1 &53.5 &44.1 &9.4 &42 \\
& FWER &34.4 &30.2 &4.3 &2.4 &43.2 &35.9 &7.4 &34\\
\cmidrule{1-10}
\multirow{2}{*}{$\tau=0.8$} & PWER &50.8 &44.8 &6.1 &3.8 &63.1 &53.1 &10.0 &52 \\
& FWER &40.4 &35.7 &4.7 & 3.0& 53.2&45.1 &8.1 &44\\
\cmidrule{1-10}
& & \multicolumn{4}{l}{$q=4/6$} & \multicolumn{4}{l}{$q=5/6$}\\
\cmidrule{1-10}
\multirow{2}{*}{$\tau=0$} & PWER &64.5 &51.3 &13.2 &5.8 &91.7 &78.7 &13.1 &92 \\
& FWER &54.6 &43.7 &10.9 &5.0 &87.2 &75.2 &12.0 &87\\
\cmidrule{1-10}
\multirow{2}{*}{$\tau=0.4$} & PWER & 71.5& 58.3&13.2 &62 & & & & \\
& FWER &61.9 &50.8 &11.1 &53 & & & &\\
\cmidrule{1-6}
\multirow{2}{*}{$\tau=0.8$} & PWER &85.8 &74.1 &11.8 &79 & & & & \\
& FWER &78.9 &68.5 &10.4 &72 & & & &\\
\cmidrule{1-6}
& & \multicolumn{4}{l}{$q=1$} & &&&\\
\cmidrule{1-6}
\multirow{2}{*}{$\tau=0$} & PWER &0 &0 &4.2 &0 & & & & \\
& FWER &0 &0 &2.2 &0 & & & &\\
\end{tabular}
\newpage
\begin{tabular}{l l l l l l l l l l}
\toprule
$l=8$& & Power & correct & false & RAE & Power & correct & false & RAE\\
\cmidrule{1-10}
& & \multicolumn{4}{l}{$q=0$} & \multicolumn{4}{l}{$q=1/8$}\\
\cmidrule{1-10}
\multirow{2}{*}{$\tau=0$} & PWER & 34.0& 34.0& 0&1.8 &35.6 &33.5 &2.2 &19 \\
& FWER &24.3 &24.3 &0 &1.3 &26.2 &24.6 &1.6 &14\\
\cmidrule{1-10}
\multirow{2}{*}{$\tau=0.4$} & PWER &34.8 &34.8 &0 &1.9 &36.9 &34.7 &2.2 &21 \\
& FWER &25.5 &25.5 &0 &1.4 &27.0 &25.4 &1.6 &15\\
\cmidrule{1-10}
\multirow{2}{*}{$\tau=0.8$} & PWER &37.4 &37.4 &0 &2.1 &40.0 &37.7 &2.3 &24 \\
& FWER &27.9 &27.9 &0 &1.6 &30.4 &28.7 &1.7 &19\\
\cmidrule{1-10}
& & \multicolumn{4}{l}{$q=2/8$} & \multicolumn{4}{l}{$q=3/8$}\\
\cmidrule{1-10}
\multirow{2}{*}{$\tau=0$} & PWER &38.0 &33.5 &4.5 &2.2 &41.3 &34.2 &7.1 &26 \\
& FWER &28.4 &25.1 &3.3 &1.7 &31.1 &25.9 &5.2 &20\\
\cmidrule{1-10}
\multirow{2}{*}{$\tau=0.4$} & PWER &39.6 &35.0 &4.6 &2.4 &43.6&36.4 &7.3 &29 \\
& FWER &29.8 &26.4 &3.4 &1.8 &33.1 &27.5 &5.5 &22\\
\cmidrule{1-10}
\multirow{2}{*}{$\tau=0.8$} & PWER &44.0 &39.0 &5.0 &2.9 &49.3 &41.4 &7.9 &35 \\
& FWER &33.6 &29.8 &3.8 &2.2 &38.4 &32.3 &6.1 &27\\
\cmidrule{1-10}
& & \multicolumn{4}{l}{$q=4/8$} & \multicolumn{4}{l}{$q=5/8$}\\
\cmidrule{1-10}
\multirow{2}{*}{$\tau=0$} & PWER &46.9 &36.9 &10.0 &3.3 &56.0 &42.3 &13.7 &45 \\
& FWER &36.7 &28.9 &7.8 &2.6 &45.9 &34.7 &11.2 &38\\
\cmidrule{1-10}
\multirow{2}{*}{$\tau=0.4$} & PWER &49.5 &39.1 &10.5 &3.6 &59.7 &45.5 &14.3 &48 \\
& FWER &39.3 &31.1 &8.2 &2.9 &49.5 &37.8 &11.7 &41\\
\cmidrule{1-10}
\multirow{2}{*}{$\tau=0.8$} & PWER &57.4 &46.0 &11.5 &4.4 &70.4 &55.3 &15.1 &59 \\
& FWER &46.6 &37.4 &9.2 &3.6 &60.9 &48.1 &12.8 &51\\
\cmidrule{1-10}
& & \multicolumn{4}{l}{$q=6/8$} & \multicolumn{4}{l}{$q=7/8$}\\
\cmidrule{1-10}
\multirow{2}{*}{$\tau=0$} & PWER &72.9 &54.6 &18.2 &6.8 &96.3 &83.8 &12.5 &96 \\
& FWER &63.8 &48.1 &15.7 &6.0 &93.4 &81.6 &11.8 &93\\
\cmidrule{1-10}
\multirow{2}{*}{$\tau=0.4$} & PWER &79.0 &61.5 &17.5 &69 & & & & \\
& FWER &70.4 &55.1 &15.3 &61 & & & &\\
\cmidrule{1-6}
\multirow{2}{*}{$\tau=0.8$} & PWER &91.3 &77.6 &13.8 &84 & & & & \\
& FWER &86.9 &74.2 &12.7 &79 & & & &\\
\cmidrule{1-6}
& & \multicolumn{4}{l}{$q=1$} & &&&\\
\cmidrule{1-6}
\multirow{2}{*}{$\tau=0$} & PWER &0 &0 &4.5 &0 & & & & \\
& FWER &0 &0 &2.3 &0 & & & &\\
\end{tabular}

\medskip
\bibliography{paperbib}
\end{document}